\documentclass[twocolumn,]{aastex631}

\usepackage{amsmath}	% Advanced maths commands
\usepackage{graphicx}	% Including figure files
\defcitealias{2006AJ....131..226Y}{YD6}
\defcitealias{bershady2010diskmass}{B10}

\begin{document}

\title[On the Stellar Disk Vertical Scale Height of Edge-on Galaxies from S$^{4}$G]{On the Stellar Disk Vertical Scale Height of Edge-on Galaxies from S$^{4}$G}

% List of institutions

\author[0009-0001-2055-644X]{Notahiana Ranaivoharimina}
\affiliation{Department of Physics and Applications, Faculty of Sciences, University of Antananarivo, Antananarivo 101, Madagascar}
\affiliation{Department of Astronomy, University of Cape Town, Private Bag X3, Rondebosch 7701, South Africa}

\author[0000-0002-8791-2138]{Toky H. Randriamampandry}
\affiliation{Valongo Observatory, Federal University of Rio de Janeiro, Ladeira de Pedro Antônio 43, Rio de Janeiro, RJ 20080090, Brazil}
\affiliation{Department of Physics \& Astronomy, University of the Western Cape, Robert Sobukwe Rd, Bellville, 7535, South Africa}

\author[0000-0002-6593-8820]{Jing Wang}

\affiliation{Kavli Institute for Astronomy \& Astrophysics, Peking University, 5 Yiheyuan Road, Haidian District, Beijing 100871, Peoople's Republic of China}
%\affiliation{Kavli Institute for Astronomy and Astrophysics, Peking University, 5 Yiheyuan Road, Haidian District, Beijing 100871, Peopleʼs Republic of China}

\author[0000-0003-3153-5123]{Karín Menéndez-Delmestre}
\affiliation{Valongo Observatory, Federal University of Rio de Janeiro, Ladeira de Pedro Antônio 43, Rio de Janeiro, RJ 20080090, Brazil}

\author[0000-0003-2374-366X]{Thiago S. Gonçalves}
\affiliation{Valongo Observatory, Federal University of Rio de Janeiro, Ladeira de Pedro Antônio 43, Rio de Janeiro, RJ 20080090, Brazil}

%% Note that the \and command from previous versions of AASTeX is now
%% depreciated in this version as it is no longer necessary. AASTeX 
%% automatically takes care of all commas and "and"s between authors names.

%% AASTeX 6.31 has the new \collaboration and \nocollaboration commands to
%% provide the collaboration status of a group of authors. These commands 
%% can be used either before or after the list of corresponding authors. The
%% argument for \collaboration is the collaboration identifier. Authors are
%% encouraged to surround collaboration identifiers with ()s. The 
%% \nocollaboration command takes no argument and exists to indicate that
%% the nearby authors are not part of surrounding collaborations.

%% Mark off the abstract in the ``abstract'' environment. 
\begin{abstract}
Disk galaxies viewed as thin planar structures resulting from the conservation of angular momentum of an initially rotating pre-galactic cloud allow merely a first-order model of galaxy formation. Still, the presence of vertically extended structures has allowed us to gather a deeper understanding of the richness in astrophysical processes (e.g., minor mergers, secular evolution) that ultimately result in the observed diversity in disk galaxies and their vertical extensions.
We measure the stellar disk scale height of 46 edge-on spiral galaxies from the Spitzer Survey of Stellar Structure in Galaxies (S$^{4}$G) project. This paper aims to investigate the radial variation of the stellar disk vertical scale height and the existence of the so-called thick disk component in our sample. The measurements were done using one-, two- and three-dimensional profile fitting techniques using simple models. We found that two-thirds of our sample shows the presence of a thick disk suggesting that these galaxies have been accreting gaseous material from its surroundings. We found an average thick-to-thin disk scale height ratio of 2.65, which agrees with previous studies. Our findings also support the disk flaring model which suggests that the vertical scale height increases with radius. We further found good correlations: between the scale height $h_{z}$ and the scale length and between $h_z$ and the optical de Vaucouleurs radius $R_{25}$. 

\end{abstract}

%% Keywords should appear after the \end{abstract} command. 
%% The AAS Journals now uses Unified Astronomy Thesaurus concepts:
%% https://astrothesaurus.org
%% You will be asked to selected these concepts during the submission process
%% but this old "keyword" functionality is maintained in case authors want
%% to include these concepts in their preprints.
\keywords{galaxies: photometry - galaxies: spiral - galaxies: vertical structure - galaxies: fundamental parameters}

\section{Introduction}\label{sec:litreview}
Previous studies have shown that knowing the vertical extent of disk galaxies allows us to derive physical and dynamical properties of galaxies such as the volumetric star formation law \citep{2019A&A...622A..64B}, the shape of the dark matter halo \citep{2010A&A...515A..63O} or the disk dynamical mass (\citealt{bershady2010diskmass}, hereafter B10). The vertical extent of galaxies can also be used to test the cosmological evolution of the universe such as the $\Lambda$ cold dark matter model \citep{2022ApJ...925..183H}. Those are not only important for our understanding of the evolution of galaxies but also the history and evolution of the universe itself. The dynamical mass density $\Sigma_\text{dyn}$ of a galaxy disk can be easily obtained using the relation $\Sigma_\text{dyn} = \sigma_z^2/\pi k G h_z$ where $\sigma_z$ is the vertical stellar velocity dispersion, $G$ the gravitational constant and $k$ a parameter depending on the nature of the disk: $k= 3/2, \ 1.71, \ 2$ for an exponential, sech and isothermal disk respectively (see \citetalias{bershady2010diskmass} for more details). 

Direct measurement of the disk thickness is only possible for galaxies that are seen edge-on or highly inclined (\citealt{2006AJ....131..226Y}, hereafter YD6). These systems allows to study the vertical extend of the stellar disk  such as the presence of thin and thick disks which have been observed in most disk galaxies including the Milky Way. \citetalias{2006AJ....131..226Y} found using R-band images that disk galaxies are well fitted with a thin and thick disk component. They also reported the thick disk component is formed by falling satellites and the thin disk is form by gas accretion. 

Numerical simulations are a useful tools to study the prevalence of thick disks and its influence on how the disk form and evolve. Recently, \citet{2023arXiv230803566Y} uses the NewHorizon simulation \citep{2021A&A...651A.109D} to study the contribution of the thick disk to the total mass of the disk and its significance. They found that the thick disk contribute more than 30 percent of the disk mass. \citet{2023MNRAS.523.6220Y} also selected Milky Way-like simulated galaxies from the FIRE-2 $\Lambda$CDM cosmological simulation \citep{2018MNRAS.480..800H} to investigate the properties of thin and thick disks in those systems.

Stellar disk flaring or the increase of the disk thickness as a function of radius have also been explored in the literature using egde-on galaxies sample. \citet{2023MNRAS.523.3915S} used the state-of-the-art cosmological simulation TNG50 and measured the disk flaring of galaxies similar to the Milky Way and the Andromeda galaxy. They found that these galaxies exhibit a wide varieties of disk flaring amplitudes and shapes. They also found that the disk scale height increases by a factor of two between the inner part to the outer part of the disk for the younger and older stellar populations \citep{2023MNRAS.523.3915S}. 
In observations, the disk flaring is measured by looking at the change in stellar height as a function of galactic radius for the Milky Way (\citealt{1998ApJ...501L..45E, 2000astro.ph..7013A}). Most studies have concluded that the disk of the Milky Way is flaring to some extend and the amplitude of the flaring depends on the stellar population (e.g. \citealt{2019ApJ...878...21T}).
Disk flaring has also been observed in several edge-on spiral galaxies (e.g \citealt{2016MNRAS.460L..89K, 2019A&A...628A..58S}).  
%These are essential for our understanding on how galaxies form and form. Our own galaxy the Milky Way is known to a have a thin disk composed mainly by old stellar population and a 

The stellar disk scale height $h_z$ of edge-on galaxies is usually measured by fitting an exponential, sech or sech$ ^2 $ models to the vertical profile (\citealt{2000A&A...361..451S}; \citealt{1997A&A...325..135X, 1999A&A...344..868X}). Isothermal distribution is more suitable for a single disk approximation and non-isothermal models (exponential and sech) correspond more to a combination of the thin and thick disks (e.g. \citealp{1991PASJ...43..755A}) which is consistent with the current observation of the Milky Way galaxy (e.g. \citealt{2018Natur.563...85H}).

%Since it is difficult to estimate the thickness of face-on and less inclined galaxies, 
Indirect methods have been adopted to predict the stellar disk thickness of face-on and less inclined galaxies using scaling relations. For instance, \citetalias{bershady2010diskmass} used the relationship between the disk oblateness $ q_{R}=h_{R}/h_{z} $ the radial-to-vertical scale length ratio and applied this relation on more face-on galaxies from their DiskMass surveys. However, the $ q_R $-$h_R$ relation has its limitation since it depends on the galaxy's morphological type and wavelength used to make these measurements (\citealt{2017A&A...605A..18C}, \citetalias{2006AJ....131..226Y}). \citetalias{2006AJ....131..226Y} found a relation between $ q_R $ and the rotation velocity derived from 21 cm observation. $ V_\text{rot} $. Other authors have also reported scaling relation between the disk oblateness and the atomic gas (\textsc{HI}) mass \citep{2005MNRAS.358..503K}. However, \citetalias{bershady2010diskmass} suggest a potential interdependence between these correlations since both rotation speed and mass are indicators of galaxy scale. Correlations between the disk oblateness and central surface brightness have also been reported in the literature (e.g. \citealt{2002A&A...389..795B, 2009ApJ...702.1567B}; \citealt{2003EAS....10..121Z}). \cite{2002A&A...389..795B} have, however, reported that this correlation yields an uncertainty of 22\% in $h_z$ measurements in the K-band. Moreover, despite using a similar sample, results from  \citet{2002A&A...389..795B} and \citet{2009ApJ...702.1567B} are not consistent with each other as some of them diverge as much as 60\%. \citet{2005AJ....130.1574S} also found that the scale height correlates with the age of the stellar population.

In this paper we explore the prevalence of thick disk by measuring the vertical extend of the stellar disk. For this, we select nearby edge-on galaxies from the  Spitzer Survey of Stellar Structure in Galaxies survey (S$^4$G; \citealt{2010PASP..122.1397S})  and measure their stellar disk thickness. We also investigate possible correlations between the disk scale height with other galaxies and compare the measured scale heights using one (1D), two (2D) and three-dimensional (3D) techniques. The near-infrared band images provided by  S$^4$G at 3.6 $\mu$m  trace the bulk of the stellar content of the disk (\citealt{2015ApJS..219....5Q}) and it is also not affected by dust obstruction. Therefore, S$^4$G allows more systematic analysis of the disk thickness compared to \citetalias{bershady2010diskmass}.

%Uncertainties in $h_z$ as calculated using equation \ref{eq:fiducial} are a combination of systematic and random errors. The random uncertainties are due to the distribution of random errors from $h_R$ measurements. Additionally, oblateness is likely to vary stochastically from one galaxy of a given scale length and type to another. Though the latter does contribute to the scatter in the fiducial relationship and introduce systematic uncertainties in the yielded value of $h_z$ for individual galaxies, it creates a random effect for the sample as a whole. Therefore, the true scale height uncertainties from the variation of disk oblateness are statistically reduced for the DiskMass survey as a whole and are considered as random errors. B10 also do not mention if the values of $h_z$ from the edge-on sample that they considered have been corrected for inclination. Further inspection of the used measurement methods leads us to infer that three of the four studies considered the inclination in their calculations. However, \citet{2002MNRAS.334..646K} and \citet{2005MNRAS.358..503K} (papers from which B10 extracted a big chunk of the results that were used to derive equation \ref{eq:fiducial}) did not take account of the inclination $i$ stating that the effect of the latter on $h_z$ would be negligible as $ i \geq 87^\circ $ for all the galaxies they studied. Although small, this could introduce more systematic error to the individual values of $h_z$.

The paper is organized as follows:
 In section \ref{chap:sample}, we present the sample and methods used to measure the scale height. 	Our results are presented and discussed in section	\ref{chap:results}. 
	In section \ref{chap:summary}, we summarise our findings and highlight possible future works. Throughout, we adopt  the AB magnitude convention, and we assume a flat $\Lambda$CDM cosmology with H$_{0}$ = 67.7 km s$^{-1}$ Mpc$^{-1}$(Planck Collaboration et al. 2016).

\section{Sample selection and methods}\label{chap:sample}
The sample selection and methods used to measure the scale height are discribed in this section. The sample selection criteria are presented and discussed in section 2.1. Section 2.2 highlights the image processing steps, theoretical background and the functional models used for the fits are described in sections 2.3 and 2.4 respectively. The possible source of uncertainties are presented in section 2.5.

\subsection{Sample selection}\label{sec:sampleselect}

We select our sample from the S$^4$G  survey \citep{2010PASP..122.1397S}. S$^4$G set out to image a volume (D $<$ 40 Mpc), magnitude (B$_{corr}$ $>$ 15.5 mag) and size (D$_{25} $ $>$ 1 arcmin) limited sample of more than 2300 nearby galaxies with the near-infrared camera (IRAC; \citealt{2004ApJS..154...10F}) in the Spitzer Space telescope at 3.6 and 4.5 $\mu$m near-infrared bands. We use 3.6 $\mu m$ band and retrieved these from the Spitzer archives from the \textsc{NASA/IPAC Infrared Science Archive} website\footnote{\url{https://irsa.ipac.caltech.edu}}. The choose of this apparent sample was motivated by the interest in having insights to the stellar distribution, free from dust obscuration and the availability of sufficiently high resolution ($\sim 1.07$ arcsec) imaging to characterize the disk scale height.

%Extinction or dust obscuration is the absorption and scattering effect of photons colliding with gas or dust particles. As a matter of fact, star light from the observed galaxies could be extincted by its own dust lane but also by the interstellar medium gas particles of our own galaxy and the dust in its plane. This effect is less prevalent in longer wavelengths (Red, infrared radiations) than in bluer light. 

%($q_0=c/a_{outer}$ $c$ is known as the intrinsic thickness)

The galaxies must be edge-on or highly inclined with an inclination larger than $ 80^\circ $. 
We note that inclination values vary within the literature. The publicly available S4G catalog available via IRSA includes Hyperleda values for inclination — in the case of NGC 4244, the Hyperleda value for inclination is 65.4 deg, while \cite{2011ApJ...729...18C} measures an inclination of 82 deg. To avoid these inconsistencies, we estimate the disk inclination i of each galaxy using Hubble’s formula:

	\begin{equation}\label{eq:incl_est}
		\cos^2 i=\dfrac{(b/a)^2-(q_0)^2}{1-q_0^2}
	\end{equation}
	where $ (b/a) $ is the disk axis ratio measured at $R_{25}$ and $q_0$ is the intrinsic flatness.  We adopted $q_0$ measured by \cite{2004ChA&A..28..127Y} using 14988 disk galaxies taken from the Lyon-Meudon Extragalactic Database. They reported a $q_0$ $\sim$ 0.11 $\pm$ 0.03 which is consistent with previous studies. The disk axis ratios were measured using the ellipse fitting technique implemented in \textsc{Photutils}, an Astropy package for source detection and photometry for astronomical sources \citep{larry_bradley_2020_4044744}.
%	As the inner ellipticity\footnote{The inner ellipticity is the ellipticity of the inner part of the disk galaxy (i.e. the bulge) which is different from the ellipticity of the disk itself (the outer ellipticity)} $\epsilon$ of disk galaxies is a function of the inner axis ratio $b/a = \cos i$ where $a$ is the inner semi-major axis, $b$ the inner semi-minor axis and $i$ the inclination, by constraining the value of 
To minimize bulge contamination, we constrain the ellipticity of the sample to  $\epsilon$ $>0.78$. In order to completely rule out irregular and early-type galaxies, we constrain the morphological type T (also known as the numerical Hubble stage T) to $ 1 \leq \text{T} \leq 8$ thus confining our data set to spiral galaxies from Sa to Sdm type.
We only include nearby galaxies with a mean distance $ D_{mean} $ less than 30 Mpc. The distances were collected from the NASA/IPAC \textsc{Extragalactic Database}\footnote{\url{http://ned.ipac.caltech.edu}} (NED) database. Distances measured using cepheid variables and the tip of the red-giant branch (TRGB) stars are preferably used since they are among the most accurate for nearby galaxies (\citealt{1993ApJ...417..553L}). We choose the distances with the smallest uncertainties for galaxies that do not have cepheid and TRGB-based distances. We note that 87\% of our sample have been measured using the Tully-Fisher method.

Considering the constraints mentioned above, the query results in a total of 65 galaxies having an ellipticity of $0.78 \leq \epsilon \leq 0.91$, a mean distance of $\ 4.1 \ \text{Mpc} \leq D_{mean} \leq 29.92 \ \text{Mpc}$ and a morphological type $3 \leq \text{T} \leq 8$ (Sb to Sdm). After inspecting the data set, 19 galaxies are dropped from the sample either because of a bright foreground star in front of the disk, a bad or noisy image or because the disk is relatively faint and hard to deblend from neighboring sources. Our final sample consists of 46 edge-on galaxies as listed in Table \ref{tab:sample}. The final sample has an ellipticity range of $0.78 \leq \epsilon \leq 0.91$, a mean distance range of $ 4.1 \ \text{Mpc} \leq D_{mean} \leq 29.23 \ \text{Mpc}$ and a morphological type range of $ 3 \leq \text{T} \leq 8$ (Sb to Sdm) essentially dominated by later types of late-type galaxies with $ 5 \leq \text{T} \leq 8$ (Sc to Sdm). 

%The closest and the farthest galaxy of our sample are as shown in Figure \ref{fig:NGC4244} and \ref{fig:IC5249}.

%\begin{longtable}{lcccccccr}\label{tab:Sample}
\begin{table*}
	\small
	%\centering
	\caption{\textbf{Properties of our sample. \textit{Column 1}: Galaxy name. \textit{Column 2}: Mean distance of the galaxy from the Milky Way taken from NED. \textit{Column 3}: Standard deviation of the distances from NED. \textit{Column 4}: Adopted distance of the galaxy retrieved from NED. \textit{Column 5}: Uncertainty on the adopted distance $ D $. \textit{Column 6}: Numerical Hubble stage or morphological type. \textit{Column 7}: Galaxy class according to the RC3 classification scheme. \textit{Column 8}: Stellar mass in $ \log_{10} M_\odot $ generated from the calibration method by \citet{2012AJ....143..139E}. \textit{Column 9}: Estimated inclination from Equation \ref{eq:incl_est}.}} 
	\label{tab:sample}
	%\endhead
	\begin{tabular}{lcccccccr}
		\hline
		\hline
		Galaxy & $ D_{mean} $ & $ D_{std} $ & $ D $ & $ \pm $ & T & Hubble class  & $  \log_{10} M_{\star} $ & $ i $ \\
		& [Mpc] & [Mpc] & [Mpc] &  & & & $  [\log_{10} M_\odot] $ & [$ ^\circ $]\\
		(1) & (2) & (3) & (4) & (5) & (6) & (7) & (8) & (9)\\
		\hline
		ESO 115-021&5.70&0.94&4.99&0.07&7.5&SBd&8.25&87.1\\
		ESO 146-014&21.4&2.36&19.4&1.79&6.5&SBcd&8.84&83.6\\
		ESO 292-014&29.5&6.05&21.2&3.81&6.7&Scd&9.81&83.5\\
		ESO 467-051&19.7&1.65&22.1&4.38&6.1&Sc&8.76&85.3\\
		ESO 482-046&23.1&3.73&22.0&1.32&5.1&Sc&9.58&84.8\\
		ESO 505-003&24.5&4.17&21.0&1.93&7.7&Sd&9.52&86.5\\
		ESO 569-014&26.5&2.63&25.4&2.34&6.4&Sc&9.77&83.6\\
		IC 0755&29.6&0.62&26.9&6.69&3.5&SBbc&9.42&83.2\\
		IC 2000&20.1&6.5&18.3&1.69&6.2&SBc&9.85&82.3\\
		IC 2058&21.5&3.7&18.1&3.25&6.5&Scd&9.5&83.8\\
		IC 3247&25.1&0.84&23.3&2.15&6.5&Sc&9.15&81.7\\
		IC 3322A&26.4&4.38&25.9&2.39&6.0&SBc&10.1&85.0\\
		IC 4213&19.6&1.53&19.9&4.12&5.8&Sc&9.21&81.9\\
		IC 5052&8.10&1.87&5.5&0.25&7.0&SBcd&9.29&81.8\\
		IC 5249&29.2&4.27&31.8&2.93&6.9&SBcd&9.43&90.0\\
		NGC 0100&16.4&3.11&18.5&1.70&5.9&Sc&9.3&82.9\\
		NGC 3044&23.3&2.25&20.2&1.86&5.6&SBc&10.3&85.3\\
		NGC 3245A&24.6&3.26&26.1&5.41&3.4&SBb&9.24&90.0\\
		NGC 3365&18.2&2.96&17.6&3.65&5.9&Sc&9.59&82.2\\
		NGC 3501&23.8&1.94&24.0&2.21&5.9&Sc&10.2&85.6\\
		NGC 4206&20.6&2.07&19.5&4.04&4.0&Sbc&9.97&84.9\\
		NGC 4244&4.10&1.74&4.29&0.14&6.0&Sc&9.12&84.1\\
		NGC 4330&20.4&0.6&18.7&3.88&6.1&Sc&9.91&85.5\\
		NGC 4437&11.6&3.06&8.34&0.81&6.0&Sc&10.2&86.9\\
		NGC 5023&10.1&3.47&6.61&0.09&5.9&Sc&9.17&82.0\\
		NGC 5348&19.8&5.19&15.9&3.29&3.8&SBbc&9.47&86.6\\
		NGC 5496&24.1&5.39&22.2&4.60&6.5&SBcd&9.7&80.8\\
		NGC 5907&16.6&1.88&16.8&0.77&5.3&SABc&10.9&90.0\\
		NGC 7064&11.4&1.27&11.6&2.30&5.1&SBc&8.95&85.1\\
		PGC 002805&16.4&0.47&18.2&3.60&6.7&Scd&8.67&80.8\\
		PGC 012798&29.3&3.86&26.7&5.29&7.6&Sd&9.63&90.0\\
		PGC 029086&10.2&2.27&14.9&2.95&7.2&Scd&8.35&81.4\\
		PGC 044358&27.7&1.79&21.8&4.32&6.7&Scd&9.95&83.8\\
		UGC 05203&28.4&1.45&32.1&6.36&6.4&Sc&9.01&84.4\\
		UGC 05245&19.0&2.05&26.4&5.23&8.0&SBd&8.63&90.0\\
		UGC 06603&24.5&2.69&25.2&4.99&5.9&SBc&9.15&80.1\\
		UGC 06667&19.4&2.62&18.3&3.62&5.9&Sc&9.31&84.0\\
		UGC 07301&21.5&3.74&25.5&5.05&6.6&Scd&8.68&81.0\\
		UGC 07321&15.0&6.25&23.1&4.57&6.5&Scd&9.19&90.0\\
		UGC 07774&19.3&6.52&27.8&5.51&6.3&Sc&9.07&81.4\\
		
%		\hline
%	\end{tabular}
%	
%\end{table*}
%\newpage
%\vspace{.5in}
%\makeatletter
%\setlength{\@fptop}{0pt}
%
%%	\pagebreak
%\begin{table*}
%	\centering
%	\small
%	\begin{tabular}{lcccccccr}
%		\hline
%		Galaxy & $ D_{mean} $ & $ D_{std} $ & $ D $ & $ \pm $ & T & Hubble class  & $  \log_{10} M_{\star} $ & $ i $ \\
%		& [Mpc] & [Mpc] & [Mpc] &  & & & $  [\log_{10} M_\odot] $ & [$ ^\circ $]\\
%		(1) & (2) & (3) & (4) & (5) & (6) & (7) & (8) & (9)\\
%		\hline
		UGC 07802&24.7&1.25&20.0&3.96&6.1&Sc&9.17&81.1\\
		UGC 07991&29.9&5.23&19.8&3.92&6.6&Scd&9.61&81.4\\
		UGC 08085&29.3&1.46&30.3&6.00&5.8&Sc&9.55&79.9\\
		UGC 09242&17.4&5.33&26.9&5.33&6.6&Scd&8.95&88.8\\
		UGC 09249&19.2&3.27&21.3&5.3&7.2&Scd&8.69&81.4\\
		UGC 09977&28.9&3.06&27.7&5.49&5.3&Sc&9.79&85.1\\
		\hline
		\hline
	\end{tabular}
	
\end{table*}
%\makeatother
%\end{longtable}

%\newpage
%\vspace{.5in}
%\begin{table*}
%	\centering
%	\small
%		\begin{tabular}{lccccr}
%			\hline
%			Galaxy  & $ D_{mean} $  & T & Hubble class  & $ \log_{10} M_{star} $& $ i $ \\
%			& [Mpc] & & & $ [M_\odot] $ & [$ ^\circ $]\\
%			(1) & (2) & (3) & (4) & (5) & (6) \\
%			\hline
%			UGC 07321&14.98&6.5&Scd&9.19&90 \\
%			UGC 07774&19.31&6.3&Sc&9.07&81.8 \\
%			UGC 07802&24.67&6.1&Sc&9.17&81.5 \\
%			UGC 07991&29.92&6.6&Scd&9.61&82.2 \\
%			UGC 08085&29.27&5.8&Sc&9.55&80.2 \\
%			UGC 09242&17.4&6.6&Scd&8.95&90 \\
%			UGC 09249&19.24&7.2&Scd&8.69&81.1 \\
%			UGC 09977&28.94&5.3&Sc&9.79&85.1 \\
%			\hline
%			\hline
%		\end{tabular}
%		
%\end{table*}

\subsection{Image processing}\label{sec:measurements}
\begin{figure*}
	\centering
	 \begin{minipage}{\columnwidth}
	 	\centering
	\includegraphics[width=\columnwidth]{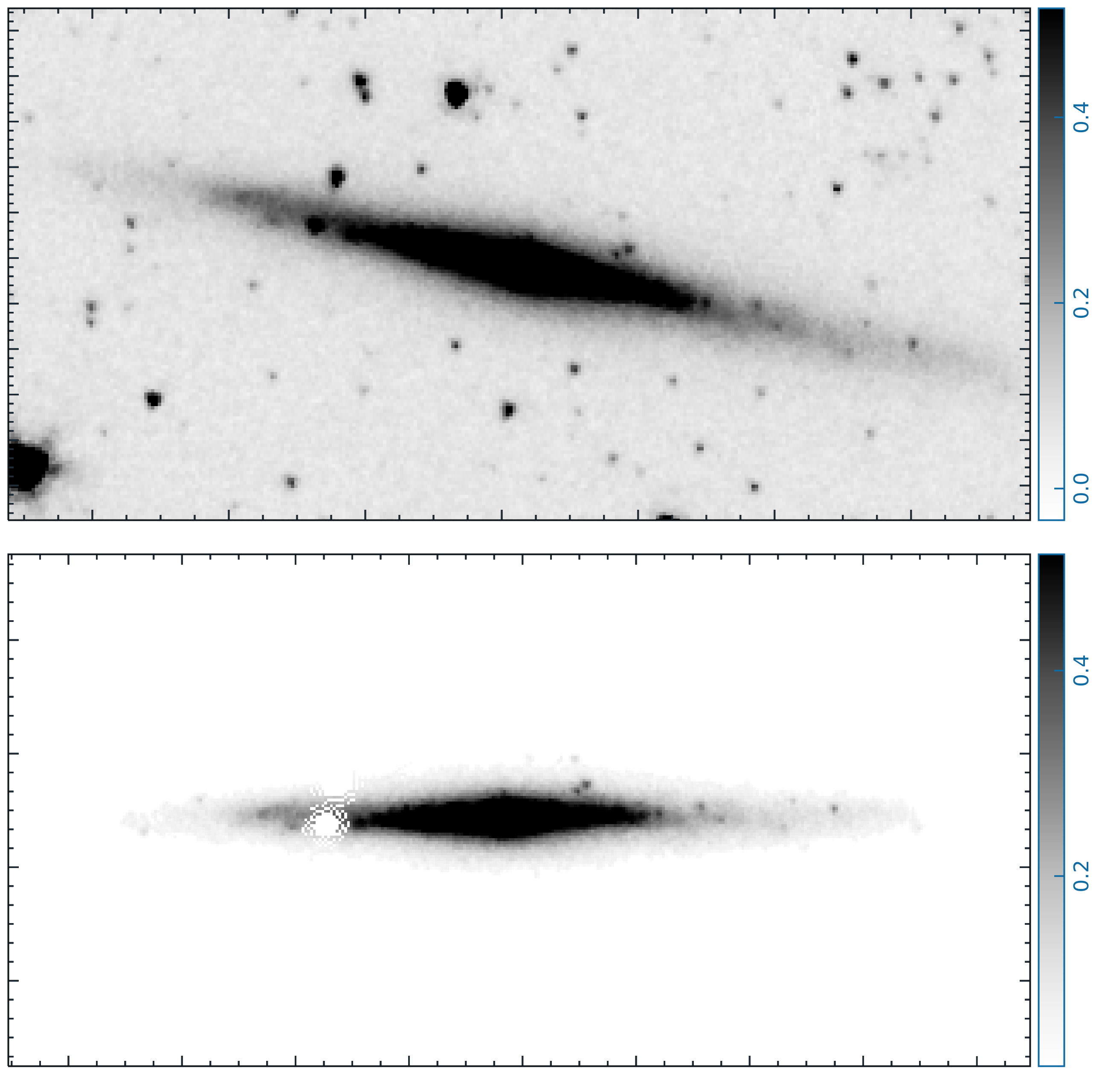}
	 \end{minipage}
	 %\vspace{0.5cm}
	 \begin{minipage}{\columnwidth}
	 	\centering
	 	\includegraphics[width=\columnwidth]{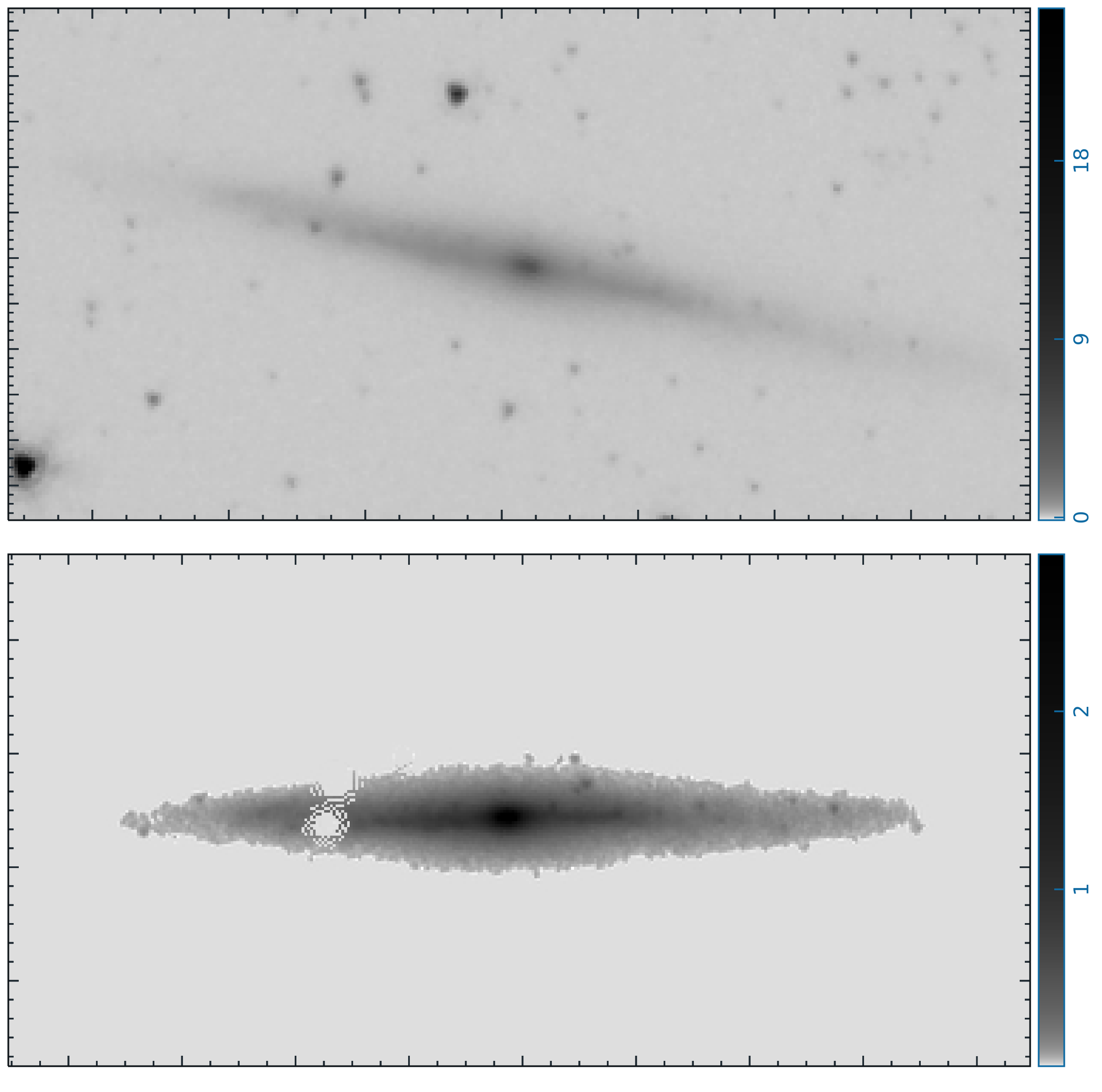}
	 \end{minipage}
	\caption{\textit{Top panels}: Uncleaned image of the UGC 09977 galaxy as seen in $3.6 \ \mu m$ band. \textit{Bottom panels}: Background subtracted, cleaned and rotated image of the above panels. A mask have also been applied to eliminate remaining foreground stars. Left panel images are scaled linearly. Right panels show log-scaled maps. All images units are in MJy/sr.}
	\label{fig:CleanedGal}
	
\end{figure*}

%\subsubsection{Background subtraction and cleaining}
Although data from S$ ^4 $G catalog have already been through basic calibration and processing, further cleaning and background subtraction are necessary before the images can be used for further analysis. We use \textsc{Photutils} \url{https://photutils.readthedocs.io/en/stable/} packages for the image processing. 
We adopt the following steps to clean the images properly.
First, we estimate the sky background and subtract by applying the Background2D technique using the \textsc{Source Extractor} algorithm implemented in \textsc{Photutils}. This algorithm uses the sigma-clipping\footnote{The sigma-clipping method is a technique in which data outliers of a sample are rejected. If the sample is to be considered to have a Gaussian distribution, deciding on which data points are to be rejected or kept depends mainly on the value of each data point's standard deviation from the sample mean $ \sigma_{clip} $. } method for which we adopted the default value of $ \sigma_{clip} = 3  $. Generally, the estimated sky background agrees with the sky value given by the S$^4 $G catalog. 
The second step is to isolate the galaxy from other foreground and background sources using image segmentation techniques. 
%Despite a successful cleaning procedure, some foreground stars or globular clusters can remain in the image and might skew the results. To go around this issue, we eliminate those using masks from the S$^4$G pipeline.
%For this, we first separate the image into multiple objects or segments (sets of pixels that have similar properties in a region). This technique is known as source segmentation.
This process can be done in \textsc{Photutils} using the \textit{detect\_source()} function. To detect sources properly, a threshold\footnote{The threshold value is a limit pixel value from which pixels are considered whether as a source or a part of the background.} value is required. This is obtained using the \textit{detect\_threshold()} function. However, source segmentation cannot set apart overlapping sources. Fortunately, the \textit{deblend\_sources()} in \textsc{Photutils} can deblend sources -- i.e. detect and separate overlapping sources into individual ones -- using a multi-threshold technique. When the data has been segmented and deblended successfully, segments not related to the galaxy (i.e background and foreground stars) are removed. 
% We use a threshold value of 3$\sigma$ where $\sigma$ is the standard deviation of the image.
The third step is to use the masks produced by the segmentation image task in \textsc{Photutils} combined with the publicly--available S$^4$G masks to properly isolate the galaxy and to minimize the contamination from remaining foreground stars and globular clusters.
%\subsubsection{Image cleaning}
%After the image data had its sky background and foreground sources subtracted, it is now possible to clean the image to "isolate" the galaxy from other foreground and background sources. 
Finally, we rotated the images so that the galaxy mid-plane should be parallel to the $ x $-axis of the image using position angle (PA) data from S$^4$G. To illustrate this, Figure  \ref{fig:CleanedGal} shows a comparison between an uncleaned image and a background-subtracted, rotated and cleaned of UGC 09977. 

% inclination angle listed in Table 1 and

%\subsubsection{Theoretical background}\label{sec:theory}

%\begin{figure}
	%\centering
%	\includegraphics[width=\columnwidth]{2Dto3D.png}
	
%	\caption{A simple sketch of an inclined disk galaxy of inclination $ i $ as seen from a plane perpendicular to the sky plane showing how a 2D image is fitted to a 3D model. $ (x_p, \ y_p) $ represents the rotated image coordinate system with $ x_p $ perpendicular to the page. $ (x_d, \ y_d, \ z_d) $ is the disk native coordinate system used for the 3D model with $ x_p  = x_d $. The variable $s$ denotes the line of sight. $ s < 0 $ points towards the Milky Way while $ s > 0 $ points away from it. \citep{2015ApJ...799..226E}} 
	%projection of the $ y_p $ component of the native disk coordinate system on the sky plane. }
%	\label{fig:2Dto3D}
%\end{figure}

\subsection{Vertical profile fitting}\label{sec:theory}
The most convenient way to measure the scale height h$_{z}$ is to fit a model to the vertical profile and retrieve the best-fit parameter values of the model. It is common to use the luminosity density profile L(z) instead of the surface brightness vertical profile. The light distributions along z of lenticular and spiral galaxies have been suggested to follow an exponential form \citep{1996A&AS..117...19D}.

\begin{equation}\label{eq:exp}
	L(z) = \mu \ \operatorname{exp}(-z/h_z) 
\end{equation}

where $\mu$ is the central density and $h_{z}$ is the scale height.
Disk galaxies having such a feature are often referred to as exponential disks. Other studies have proposed an isothermal disk model for light distribution profile (e.g van der Kruit and Searle 1981): 
\begin{equation}\label{eq:sech2}
	L(z) = L_0 \ \operatorname{sech}^2(z/z_0) 
\end{equation}
where $ L_0 $ is the luminosity density at galactic mid-plane and $z_0$ is a vertical scale parameter such as $z_0 = 2h_z$. In this case, the constant scale parameter $z_0$ at all $ z $'s is given as \citep{1981A&A....95..105V}:
%On the other hand, $z_0$ is dependent on the vertical velocity dispersion $ \langle{V^2_z}\rangle^{1/2} $  \citep{1981A&A....95..105V}:

\begin{equation}\label{eq:z0veldisp}
	z_0 = \dfrac{\langle{\sigma^2_z}\rangle^{1/2}}{(2 \pi G\varrho_0)^{1/2}}
\end{equation}
%At the same time, $ \langle{V^2_z}\rangle^{1/2} \propto T$ (here $ T $ is the temperature). This implies that if a disk is isothermal (i.e. $ T $ is constant at all $ z $) thus $z_0$ is also constant
where $G = 6.67 \times 10^{-11} \ \text{N m}^2 \ \text{kg}^{-2}$ is the gravitational constant and $ \varrho_0 $ the space density at the mid-plane. \citet{1988A&A...192..117V} also proposed the following model to fit the vertical profile:
\begin{equation}
	L(z)= L_0 \ \operatorname{sech}(z/h_z)
\end{equation}
These functions have similar behavior at large-scale heights. However, the non-isothermal profiles are suitable for galaxies with thick and thin disk components which have non-continuous potentials at mid-plane (B10 and references therein). 
%At large heights, the two models look quite similar. At intermediate values of $ z $ the sech$ ^2 $ have been proven to fit the profile better compared with the simple exponential function. Though, on very low heights, the sech$^2$ model fails to fit the vertical profile. Note that this region near the mid-plane is where temperatures are the lowest in disk galaxies. This also is where most of star formation activity occurs and where the galactic dust lane is prominent. It is then mandatory to reject the isothermal assumption and thus the sech$ ^2 $ within this region. In this context, 
%Regardless, these models are usually the simplest and most used for one-dimensional (1D) fitting.

Additionally, \citet{1981A&A....95..105V} also proposed a two-dimensional approach to measure the scale height. Their model allows to fit the surface brightness along the radius and the vertical axis simultaneously giving the best-fit parameters values of the scale length and the scale height. It is the same model as the EdgeOnDisk model of the data analysis software \textsc{Galfit} \citep{2010AJ....139.2097P}. Assuming an isothermal disk, the proposed model is:
% \begin{equation}\label{eq:vanderKruit}
% 	\mu (R, z) = \mu (0, 0) \ (R/h_R) K_1 (R/h_R) \ \operatorname{sech}^2(z/z_0) 
% \end{equation}
\begin{equation}\label{eq:vanderKruit}
	\mu_\text{Jy} (R, z) = \mu_\text{Jy}(0,0) (R/h_R) \ K_1 (R/h_R) \ \operatorname{sech}^2(z/h_z)
\end{equation}
% \mu (0, 0) = 2 \ h_R L_0 
with $ \mu_\text{Jy}(0,0) $ the central surface brightness and $ K_1 $ the modified Bessel function. A later study by \citetalias{2006AJ....131..226Y} proposed a generalized version of Equation \ref{eq:vanderKruit}. The latter allows additional flexibility by adding a new parameter, $n$, that permits a variation in the profile shape (i.e. the function that the profile follows). Here the surface brightness has units of flux density (usually in Jansky\footnote{1 Jy = 10$ ^{-26} $ W m$ ^{-2} $} or Jy): 
\begin{equation}\label{eq:YD}
	\mu_{\rm Jy} (R, z) = \mu_{\rm Jy} (0, 0) \ (R/h_R) K_1 (R/h_R) \ f(z) 
\end{equation}
where f(z) = $\operatorname{sech}^{2/n}(nz/z_0)$ and $K_1$ the modified Bessel function. When $ n \rightarrow 1 $, Equation \ref{eq:YD} becomes Equation \ref{eq:vanderKruit} which is an isothermal disk having a sech$ ^2 $ shape. Obviously, $ n \rightarrow 2 $ has the vertical component of (\ref{eq:YD}) that follow the sech function model.  In contrast, with $ n \rightarrow \infty $, the last block of relation \ref{eq:YD} becomes $ f(z) \propto \exp(-z/h_z) $. We now have an exponential disk of scale height $h_z$. 

%Converting the surface brightness $ \Sigma $ to $ \mu $ can be easily done using the relation:
%\begin{equation}\label{eq:fluxtomag}
%	\mu = \mu_{ZP} - 2.5 \log_{10}{\Sigma}
%\end{equation} 
%where $ \mu_{ZP} $ is the photometric zero point (ZP) magnitude. Practically, the ZP magnitude is a value used to convert instrumental magnitudes into a standard magnitude system.

Both Equations \ref{eq:vanderKruit} and \ref{eq:YD} are exclusive to perfectly edge-on galaxies (i.e. $ i = 90^\circ $) since they do not consider the effect of inclination of highly inclined galaxies (i.e. $ 80^\circ \leq i  < 90^\circ $). To tackle this issue, 3D models have been proposed (e.g. \citealt{2015ApJ...799..226E}). These 3D models are essentially based on 2D models that, instead of computing pixels of a 2D image, they are computed on a native galactic disk three-axis coordinate system ($xd$, $yd$, $zd$), where these axes are orthogonal to the galaxy's semi-major axis of the disk, respectively \citep{2015ApJ...799..226E}; Figure 7 from \citet{2015ApJ...799..226E}  illustrates this approach for an inclined spiral galaxy, the transformation from a two-axis image coordinate system ($ x, \ y $) to the galactic disk there-axis coordinates uses the inclination $ i $ and takes the line of sight $s$ into account. It is common to use cylindrical coordinates instead of galactic disk coordinates with $R=(x_d^2 + y_d^2)^{1/2}$ is and $z=z_d$. 
The generalized model in terms of luminosity density is:
\begin{equation}
	L(R,z)=L_0 \ \exp(-R/h_R) \ \operatorname{sech}^{2/n}(nz/z_0)
\end{equation}

%The surface brightness distribution of disk galaxies is observed as having a decreasing trend from its mid-plane to higher $ z $'s. The variation of the surface brightness along the vertical axis is known as the vertical profile $ \mu(z) $. The most convenient way to measure the scale height $h_z$  is to fit a model to the vertical profile and retrieve the best-fit parameter values of the model. Theoretically, it is common to use the luminosity density profile $ L(z) $ instead of the surface brightness vertical profile which is the luminosity density distribution along the vertical axis. Originally, light distributions along $z$ of lenticular and spirals were suggested to follow an exponential law i.e. $ L(z) \propto \exp{(-z/h_z)} $ \citep{1996A&AS..117...19D}. Disk galaxies having such a feature are often referred to as exponential disks. Assuming a symmetrical and isothermal disk, later studies proposed the following model for light distribution profile:

%\paragraph{Zero point and PSF}
%Before any scale height measurement is performed, it is crucial to estimate the instrument’s zero point to ensure accurate results. 

%In reality, observed surface brightnesses are measured in native instrumental magnitudes 

\subsection{Fitting procedures}
We measure the scale height by fitting the vertical profile to a known model. The pixel values of images in units of flux density (MJy per steradian\footnote{1 MJy = $10^6 $ Jy; 1 steradian = $ 4.25 \times 10^{10} \ \text{arcsec}^2 $} for the S$ ^4 $G images) have to be converted into units of surface brightness (mag arcsec$^{-2}$). For the 3.6 $ \mu $m band of the S$ ^4 $G survey, the zero point flux density in Jansky is 280.9 Jy \citep{2008AJ....136.2761O}. Basically, the conversion from MJy/sr to mag arcsec$^{-2}$ for the 3.6 $ \mu $m is given by the relation:
\begin{equation}\label{eq:3.6umFluxtoMag}
	\mu_{3.6 \ \mu\text{m}} = -2.5 \times \log_{10}\left[\frac{{S_{3.6 \ \mu\text{m}}} \times 2.35 \times 10^{-5} }{280.9}\right]
\end{equation}  
with $ S_{3.6 \ \mu\text{m}} $ the pixel value in MJy/sr.

%To mathematically reflect the nature of photometric observation, we convolve a point spread function (PSF) to our data. To smooth and remove outliers in our data, we convolve the image with a synthetic PSF. 
To smooth and remove outliers in our data, we first generate a 2D Gaussian function of FWHM = 2.2 arcsec using the \textsc{Astropy.convolution} attribute \textit{Gaussian2DKernel} (\citealp{astropy:2013}, \citeyear{astropy:2018}) and convolve the latter with the background-subtracted, cleaned and rotated images. In the following, we introduce the fitting procedure for the 1D, 2D and 3D methods:
%For the 3.6 $ \mu $m band of the S$^4$G survey, the PSF is equivalent to a Gaussian with a full width half maximum (FWHM) of 2.2 arcsec (\citealt{2010PASP..122.1397S}, \citealt{2011ApJ...741...28C}). 
% the IRAC-1 point spread function (PSF)
%\paragraph{1D method}
%\begin{enumerate}

 We first use the 1D method for two-component disk investigation as the existence of such a feature can be easily noticeable by observing the profile. We then average the values of pixels along the image's x-axis. This allows having an averaged vertical profile that is converted into surface brightness units using Equation \ref{eq:3.6umFluxtoMag}. Next, we convolve the model with a 1D Gaussian kernel with an FWHM of 2.1 arcsec since the images were previously smoothed. Here we use the sech$^2$ and exponential functions as models considering the vertical profile shape at extreme cases ($n \rightarrow 1, \ N \rightarrow \infty $ respectively, see Equation \ref{eq:YD}). We fit our image to a two-component model of each of the two mentioned above using \textsc{lmfit} \citep{2016ascl.soft06014N}. This Python package includes a large number of methods for curve fitting. We choose the Least-Squares algorithm returning a reduced chi-squared value $\chi^2_\nu$ when the fit is done. This parameter gives an insight into the quality of fits. Theoretically, a good fit has $\chi^2_\nu = 1$. However, a non-biased value of this parameter requires the data to be weighted properly. We use the variance map of the images as weights for our data. We consider that a galaxy shows the presence of a thick disk when the two-component fit is esteemed to be successful i.e. the resulting scale height relative uncertainty of each component is lower than 30\% and $\chi^2_\nu \sim 1$. In the case $\delta h_{z \ thin}/h_{z \ thin} < 30\%$, $\delta h_{z \ thick}/h_{z \ thick} \geq 30\%$ and $\chi^2_\nu \sim 1$, we treat the galactic disk as a single-component disk. Else, we consider the fit unsuccessful.   

%\begin{equation}
%\chi^2_\nu = \frac{1}{\nu} \ \sum_{i}  \frac{(f_{\text{data}}(x)-f_{\text{model}}(x))^2}{\sigma_i^2}
%\end{equation}
%with $ \nu $ the number of degrees of freedom,

%In a previous study, \citeauthor{2007ApJ...662..335B} showed that the vertical profile shape approaches the sech$^2$ as the fits are performed on increasing radii. Hence, because we fit over an the entire diameter of the galaxy, we expect the the data to be as shallow as the sech$^2$ model.

We also employ the 1D method box model approach as a way to track the variation of the scale height along the radius giving us the radial profile of the scale height. We retrieve the vertical profiles along the radius using a similar approach as the ``BoxModel'' task within the \textsc{NOD3} program \citep{2017A&A...606A..41M}. We divide the galaxy into small boxes instead of averaging along the entire length of the galaxy. Pixel values within each box are then averaged along the y-axis. This enables us to obtain the vertical profiles along the radial extent of the galaxy that can be fitted to a model. We fit the profile using a single component sech or exponential functions. We do not use the sech$^2$ model here since previous papers have shown that vertical profiles are steeper than the sech$^2$ model when observed from specific radii \citep{2007ApJ...662..335B}.  
%\paragraph{2D method}
% \item We handle the 2D approach using the most commonly used and extensively tested data analysis software \textsc{Galfit} \citep{2010AJ....139.2097P} instead of python packages. This program is widely used for galaxy fitting using the 2D method. The \textsc{Galfit} model for edge-on galaxies (the EdgeOnDisk model)takes on the form of Equation \ref{eq:vanderKruit} except that it uses the vertical scale height $h_z$ instead of the scale parameter $z_0 $:
 We handle the 2D approach using the most commonly used and extensively tested data analysis software \textsc{Galfit} \citep{2010AJ....139.2097P} instead of python packages. This program is widely used for galaxy fitting using the 2D method. As mentioned previously, the \textsc{Galfit} model for edge-on galaxies (the EdgeOnDisk model) takes on the form of Equation \ref{eq:vanderKruit}. \textsc{Galfit} uses a Least-Squares algorithm that uses Levenberg–Marquardt technique to perform the fits yielding a value of $\chi^2_\nu$:
\begin{equation}
	\chi^2_\nu = \frac{1}{\nu} \ \sum_{x=1}^{Nx} \sum_{y=1}^{Ny} \frac{[f_{\text{data}}(x,y)-f_{\text{model}}(x,y)]^2}{\sigma(x,y)^2},
\end{equation}
where $\nu$ is the number of degrees of freedom, $ Nx $ and $ Ny $ the image dimensions, $ f_{\text{data}}(x,y) $ the flux of the $ (x, y) $-pixel, $ f_{\text{model}}(x,y) $ the sum of $ M $ functions of $ f_i(x, y; \ \alpha_1, \ ..., \alpha_n) $, with $ N $ free parameters $ (\alpha_1, \ ..., \alpha_n) $  in the above model and $ \sigma(x,y) $ the Poisson error that requires the image gain and readout noise\footnote{The gain and readout noise are quantities that characterizes the image sensors used in a telescope. Gain is the amount of amplification as the detected charges are converted into voltages. Readout noise is the noise of the amplifier.} information as weights. These can be found in the image header file. 
Input images need to be in units of count. Therefore, we convert pixel values of the image from MJy/sr to counts by multiplying each pixel value with the exposure time and dividing it by the flux conversion factor. We choose to fit images using a single component edge-on-disk model for simplicity and to minimize computing time. We keep the background value fixed while the PA and galaxy center parameters are left to vary during the fitting process.

%\paragraph{3D method}
 Because \textsc{Galfit} is essentially software for 2D fitting, we perform 3D method fits using \textsc{Imfit}, a program for astronomical image fitting \citep{2015ApJ...799..226E}. In this regard, we use the ExponentialDisk3D function derived from Equation \ref{eq:YD} as a model. The latter utilizes the same principle as explained in Section \ref{sec:theory} to fit the 2D image onto the 3D model. Among the different minimization algorithms and statistics that \textsc{Imfit} has, we adopt the Levenberg–Marquardt technique coupled with the Poisson maximum-likelihood ratio statistic as recommended by  \citep{2015ApJ...799..226E} to make sure fits are accurate and performed as fast as possible. 

%To prevent the fits from being computationally expensive, we use a single-component ExponentialDisk3D model and fix the value of the inclination parameter to the inclination angle from the axis ratio. Nonetheless, we let the $n$ parameter that dictates the profile shape vary.
%However, this causes $ \chi_\nu^2 $ to be unreliable as a parameter for quality-of-fits description.
%\end{enumerate}

\subsection{Uncertainties}
\subsubsection{Inclination correction}\label{sec:incl_corr}
\begin{figure*}
	\vspace{-0.5cm}
	\centering

    \begin{minipage}[b]{0.475\textwidth}
        \centering
	   \includegraphics[width=\textwidth]{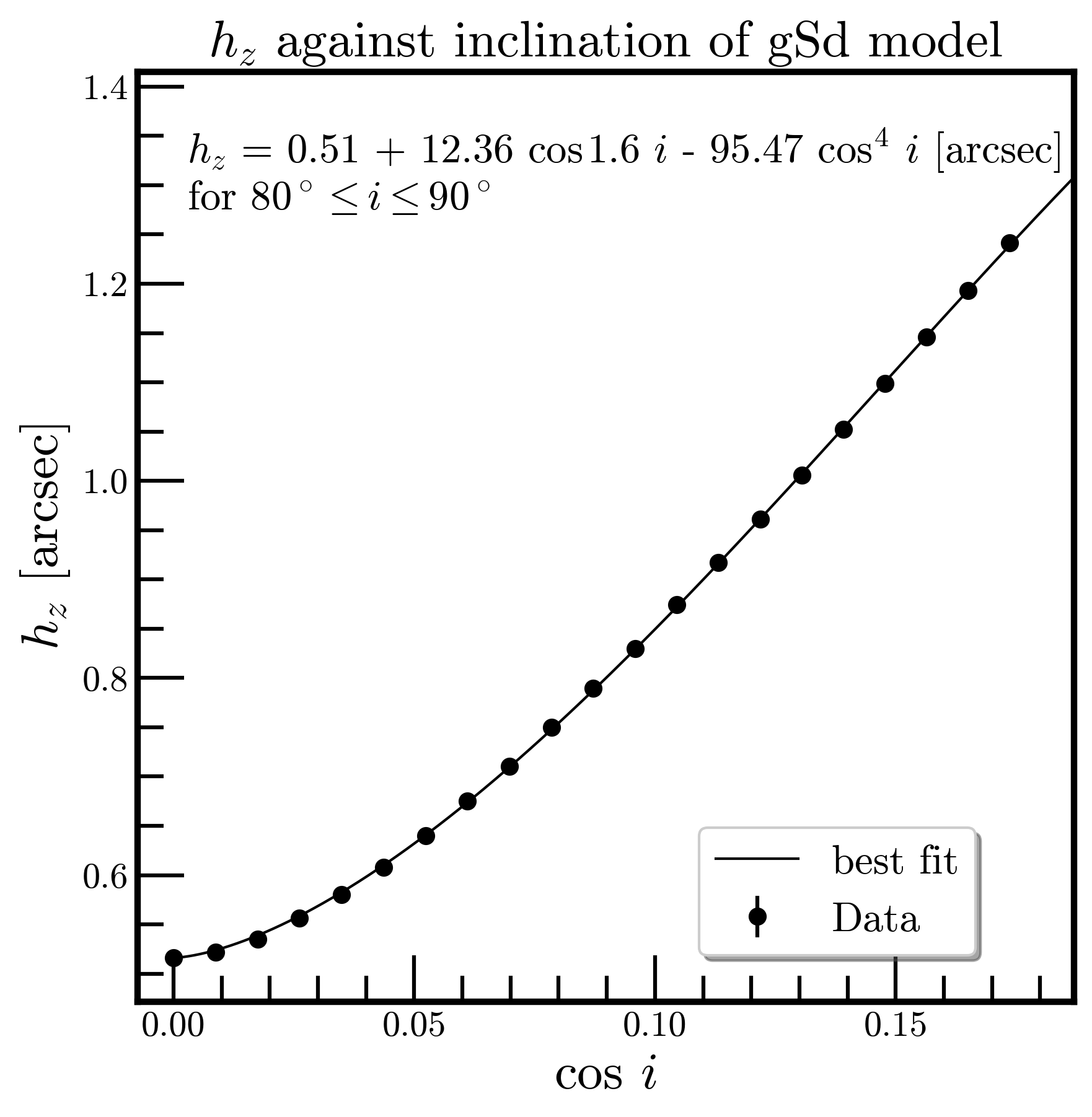}
    \end{minipage}
        \hfill
    \begin{minipage}[b]{0.5\textwidth}
        \centering
        \includegraphics[width=\textwidth]{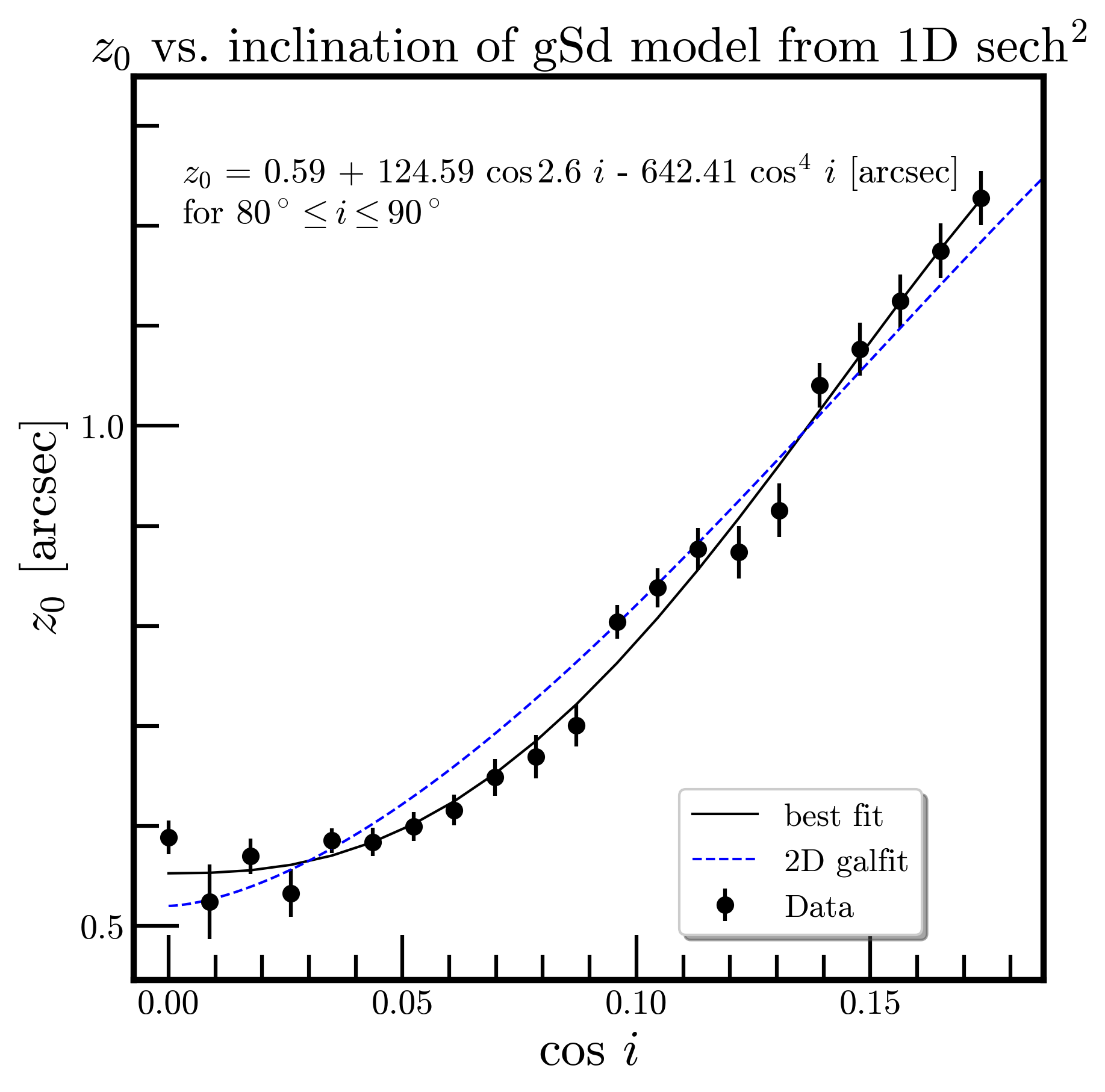}
    \end{minipage}
	\caption{Variation of the scale height $h_z$ and vertical scale parameter $z_0=2h_z$ against inclination $i$ obtained from different inclination angles of the gSd model of the GalMer database. \textit{Left Panel: }Values of $h_z$ derived using the 2D method. \textit{Right Panel: }Values of $z_0$ obtained using the 1D method compared to those from the Galfit 2D method (solid blue line). The ensuing relation can be used for inclination correction for results from both the 1D and the 2D methods for galaxies of inclination $80^\circ \leq i \leq 90^\circ$.}
	\label{fig:HzvsIncl}
\end{figure*}
Since the 1D and 2D methods do not take into account the galaxy inclination, therefore, the measured scale height results can be over/under-estimated if the galaxy is not perfectly edge-on. 
%Known direct inclination correction relations have been obtained from galaxies of inclination $ i \leq 73^\circ $. Correction for edge-on galaxies have been done by extrapolation from the mentioned relations or using other indirect methods. Regardless, any technique might introduce bias for highly inclined disk galaxies. 
For this study, we use the online simulation GalMer \citep{2010A&A...518A..61C}. It is mainly used for galaxy mergers and interaction simulations. The GalMer database has a wide range of galaxy models with different types of morphology that can be rotated and manipulated. To estimate the uncertainties associated to these models as a function of inclination, we choose a non-interacting giant Sd-type spiral model (late-type spiral) known as the gSd model with no companion (stars only). To change the inclination, we vary the angle $ \theta = 90^\circ - i$ from 0$ ^\circ $ to 10$ ^\circ $ in 0.5$ ^\circ $ increments in order to be consistent with our sample. We then download each image of the model emanating from each value of $ \theta $. After that, images are fitted using the 2D method yielding the variation scale height $h_z$ as a function of inclination $i$ (left panel Figure \ref{fig:HzvsIncl}). The resulting relation for the gSd model is:

\begin{equation}\label{eq:inclcorrection}
	h_{z,edge-on} = h_{z,measured} - A \ \cos^{1.6} i - B \ \cos^4 i \ \ [\text{arcsec}] 
\end{equation}
where $h_{z,edge-on} = 0.51$ is the scale height for a perfectly edge-on galaxy model, $A$ and $B$ are real coefficients equal to 12.36 and 95.47 respectively with $ 80^\circ \leq i \leq 90^\circ $ only. We also perform the same calibration using the 1D method. A comparison between the relation obtained from the 1D and 2D methods is shown in the right panel of Figure \ref{fig:HzvsIncl}. We observe that both correlations are equivalent although the one from 2D GALFIT seems to be more accurate since data point uncertainties are smaller. Hence, scale height results of galaxies having a measured inclination $i$ from both 1D and 2D methods are corrected using the relation given by Equation \ref{eq:inclcorrection}. Based on testing on galaxies with known inclinations, we expect that values of $h_z$ obtained from 1D and 2D methods and corrected using our relation agree at within 10\% with $h_z$ results derived from the 3D method.

Not accounting for inclination may also underestimate the central surface brightness results. To correct this, we employ a relation given by \citet{1970ApJ...160..811F} using J2000 thus dropping the cosecant term :
\begin{equation}\label{eq:mu0inclcorrection}
	\mu(0)_c = \mu(0) + 2.5 \ \log_{10} \frac{a}{b}
\end{equation}
where $ \mu(0)_c $ is the corrected central surface brightness, $ \mu(0)_c $ is the observed central surface brightness, $ \frac{a}{b} $ the ratio of major-to-minor axis of the disk. We note that this correction assumes that the disk is infinitely thin. In fact, our measurements, presented in Section \ref{chap:results}, show an average disk scale height $\leq$ 0.5 kpc supporting that the analyzed disks are indeed thin.

\subsubsection{Systematic uncertainties}\label{sec:syst_err}
% As previously mentioned in Section \ref{sec:sampleselect}, we adopted distances issued from the NED database from different papers using different values of $H_0$ and redshift-independent distance ladder methods (essentially the Tip of the Red Giant Branch or TRGB method and the Tully-Fisher method). Those distances contain uncertainties that can introduce errors in scale height results while used for converting from units of arcseconds to kiloparsecs.

As previously mentioned in Section \ref{sec:sampleselect}, we adopted distances issued from the NED database. One needs to correct and unify the distances due to different values of $H_0$ and redshift-independent distance ladder methods (essentially the TRGB and the Tully-Fisher methods). Those distances contain uncertainties that can introduce errors in scale height results while used for converting from units of arcseconds to kpc.

Galaxies not exactly having a PA of $ 90^\circ $ after image cleaning can also be a source of error for results obtained using the 1D method. This type of uncertainty is expected to be included in statistical errors yielded from individual fit results after correction using the method calibrated in Section \ref{sec:incl_corr}. 

%Even though we deem masks from the S$^4$G pipeline trustworthy, some profile-contaminating foreground stars might be leftover introducing more uncertainties to our results. On top of this, masking  

%(all within 1.8 \%). Used values of $ H_0 $ happens to fall within $ 74 \pm 1$ km s$^{-1}$ Mpc$^{-1}$.

\section{Results and discussion}\label{chap:results}
Comparisons between the measured scale height using 1D, 2D and 3D techniques are given in Section \ref{sec:modelcomparison}, as well as a discussion on the one-component disk model. The two-component disk model results are shown in Section \ref{sec:res2comp}. In Section \ref{sec:diskflaring}, we analyze the radial profile of the scale height. We present our findings regarding the correlations between the scale height and other global galaxy properties in Section \ref{sec:correlations}.

\subsection{Model comparison}\label{sec:modelcomparison}
%Although the average $h_z$ results for all measurement methods should be in agreement
 In Table \ref{tab:Modelcomparison} we summarize the measured scale height derived from 1D, 2D and 3D methods. The average uncertainties are the total errors i.e. the combination of statistical and systematic errors. We find that 2D and 3D methods produce the smallest uncertainties and are typically in good agreement with each other. The number of galaxies $N$ for which the fits were successful are given in column 2 in Table \ref{tab:Modelcomparison}. We choose the model with a high success rate and small uncertainties for the statistical analysis part of this paper. This leads us to consider only the results from the 2D method for the one-component disk analysis. The 1D sech$ ^2 $ and the sech box model results are used for the two-component disks analysis and the radial profile of the scale height, respectively.

\begin{table*}
	
	\caption{Summary of $h_z$ results from each model. \textit{Column 1:} Model name. \textit{Column 2:} Number of galaxies that had a successful fit. \textit{Column 3:} Mean inclination-corrected scale height $h_z$. \textit{Column 4:} Standard deviation of the $h_z$. \textit{Column 5:} Average relative uncertainty on $h_z$. \textit{Column 6:} Standard deviation of the relative uncertainty on $h_z$.}
	\label{tab:Modelcomparison}
	\small
	%\centering
	\begin{tabular}{lccccc}
		\hline
		\hline
		Model &  $ N $ & Mean $h_z$ & Std $h_z$ & Average $\delta h_z/h_z$ & Std $\delta h_z/h_z$ \\
		& & [kpc] & [kpc] & [\%] & [\%] \\
		(1) & (2) & (3) & (4) & (5) & (6)  \\
		\hline
		1D sech$^2$ thin disk&43&0.14&0.07&19.0&6.87\\
		1D sech$^2$ thick disk&29&0.34&0.16&17.1&7.90\\
		1D exponential thin disk&9&0.20&0.05&19.9&7.01\\
		1D exponential thick disk&7&0.51&0.27&20.0&8.73\\
		1D box model exponential &46&0.47&0.22&27.4&9.17\\
		1D box model sech &46&0.40&0.18&23.6&7.39\\
		2D \textsc{Galfit} EdgeOnDisk &46&0.33&0.14&15.4&6.69\\
		\textsc{Imfit} ExponentialDisk3D &30&0.32&0.12&14.9&6.70 \\
		\hline
		\hline		 
	\end{tabular}
\end{table*}

%George & Mallery, 2010 (Pearson skewness coefficient)

\begin{figure}
%	\vspace{0cm}
		%\centering
		\begin{minipage}[b]{0.475\textwidth}
	\centering
	\includegraphics[width=\textwidth]{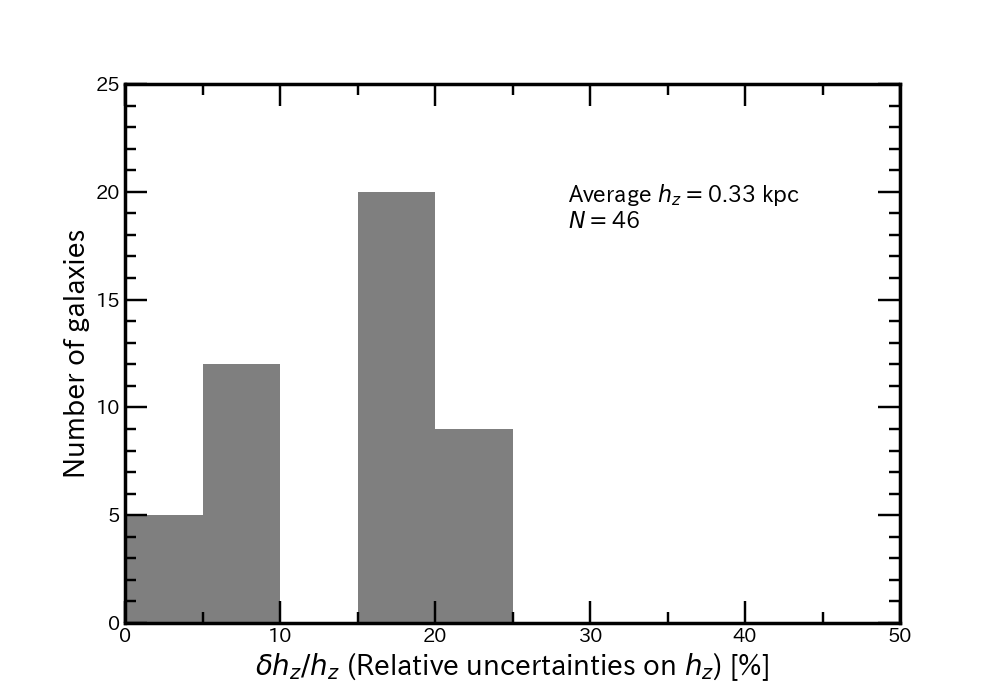}
	%\caption[Histogram of result uncertainties from 2D EdgeOnDisk model]{Histogram of result uncertainties from 2D EdgeOnDisk model}
	%	\end{minipage}
	%\label{fig:histerr2D}
	%		
		\end{minipage}
%\begin{figure}
    \hfill
	\begin{minipage}[b]{0.475\textwidth}
%	\vspace{-0.5cm}
	\centering
	\includegraphics[width=\textwidth]{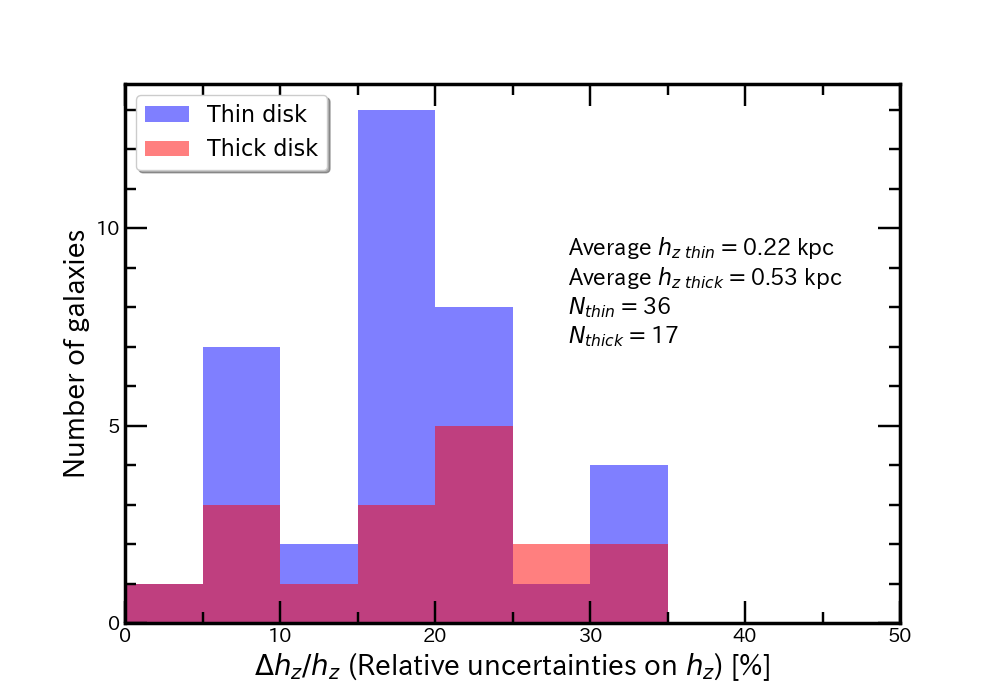}			
	\end{minipage}
 \caption{\textit{Top Panel: }Histogram of resulting uncertainties from the 2D EdgeOnDisk model. \textit{Bottom Panel: }Histogram of resulting uncertainties from the 1D sech$^2$ model}
	\label{fig:histerr}
\end{figure}

The histogram of the uncertainties for the 2D model is shown in the top panel of Figure \ref{fig:histerr}. It gives the percentage of the uncertainties distribution relative to the measured scale height. The uncertainties distribution resembles a left-skewed normal distribution at first glance with the exception of a gap in the middle. However, its Pearson skewness coefficient is estimated to be -1.96 which falls into the acceptable range in order to prove normal distribution. The presence of the gap most likely originates from the strong influence of galaxy distance errors on $h_z$ uncertainties (see Section \ref{sec:syst_err}). In fact, the distance measured using the TRGB technique yields less uncertainty than the Tully-Fisher distance. Therefore, the two different distance measurement may have introduced the gap in the left panel of Figure \ref{fig:histerr}. However, due to the presence of a random error as discussed in Section \ref{sec:syst_err} the gap is not observed in the histograms of errors for 1D models and they instead show a bimodal trend. The histogram of uncertainties for 1D sech$ ^2 $ model is shown in the bottom panel of Figure \ref{fig:histerr}. 
%The skewness coefficients for the thin and thick disk is equal to -1 and -1.46 respectively. 

\begin{figure*}
	%	\hspace{-2cm}
	%\centering
	\begin{minipage}{0.45\textwidth}
		\centering
		\includegraphics[width=1\textwidth]{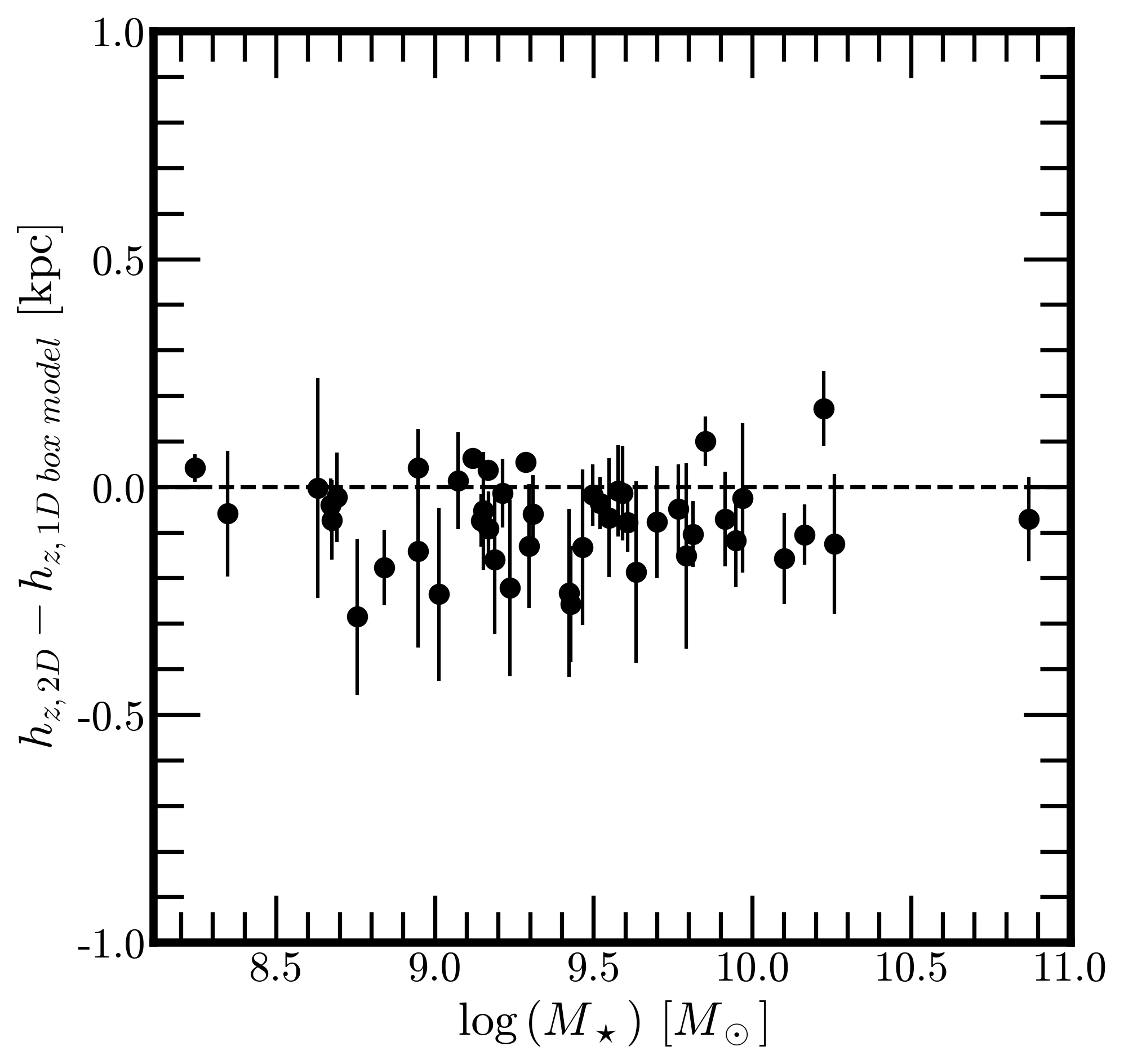}
	\end{minipage}
	\hfill
	\begin{minipage}{0.45\textwidth}
		\centering
		\includegraphics[width=1\textwidth]{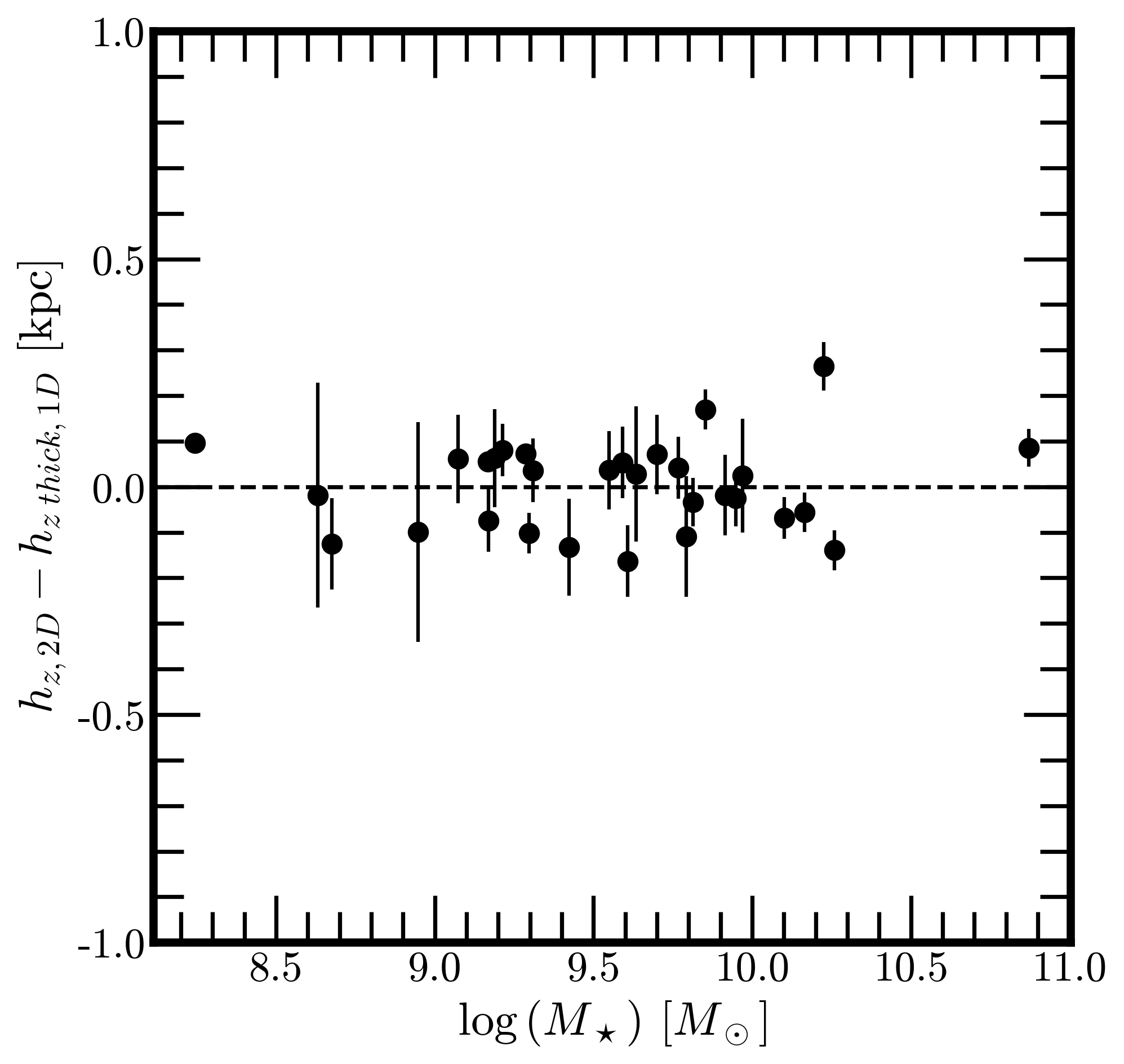}
	\end{minipage}
	%		\hfill
	%		\begin{minipage}{0.32\textwidth}
		%			\centering
		%			\includegraphics[width=1\textwidth]{1D_sech_thick_disk_vs_corrected_galfit_2D_best_var_weights.png}
		%		\end{minipage}
	\caption{Comparison of scale height results from 1D method and 2D method. \textit{Left Panel:} The average $h_z$ from 1D sech box model and $h_z$ from \textsc{Galfit} EdgeOnDisk model. \textit{Right Panel:} Thick disk $h_z$ from 1D sech$^2$ model and $h_z$ from 2D EdgeOnDisk model.} %\textit{Right Panel:} Thick disk $h_z$ from 1D sech model and $h_z$ from 2D EdgeOnDisk model.}
\label{fig:1Dvs2D}
\end{figure*}

Figure \ref{fig:1Dvs2D} compares the scale heights results from the 1D models to $h_z$ to the 2D models. The average h$_{z}$ from the box sech model and the 2D sech$^{2}$ is plotted on the left panel. Right panel shows the thick disk scale height from the 2 disk sech$^2$ model as a function for the 2D scale height. The left panel of Figure \ref{fig:1Dvs2D} suggests that when flaring is considered the 1D sech model has systematically slightly higher $h_z$ than the 2D sech$^2$ model. It also implies that the 2D sech$^2$ model may slightly miss the flaring outer part of the disk, if we assume the systematic difference in $h_z$ between the sech$^2$ and sech models can be ignored since both models converge to the exponential model at large radius. However, the two measurements are consistent with their uncertainties. The right panel of this figure shows that the 2D sech$^2$ single-component measurements are consistent with the measurements of the thick component of the 1D sech$^2$ two-component model. The results of the right panel suggest that the 2D sech$^2$ model tends to measure the relatively thicker disk component even if a thin disk exists.

We present the best-fit parameters for the single component disk in Table \ref{tab:2Dresults}. While uncertainties on h$_{z}$ and h$_{R}$ are total errors, uncertainties on the central surface brightness are statistical errors only. Both scale heights and central surface brightness are corrected for inclination using the model given by Equations \ref{eq:inclcorrection} and \ref{eq:mu0inclcorrection} respectively.
%Both parameters are comparable to those obtained from studies on face-on galaxies.
%Radial scale lengths were not corrected for inclination as they are not affected by the angle at which the galaxy is seen from our perspective.

\begin{figure}
	%	\vspace{-.25cm}
	\centering
	\includegraphics[width=\columnwidth]{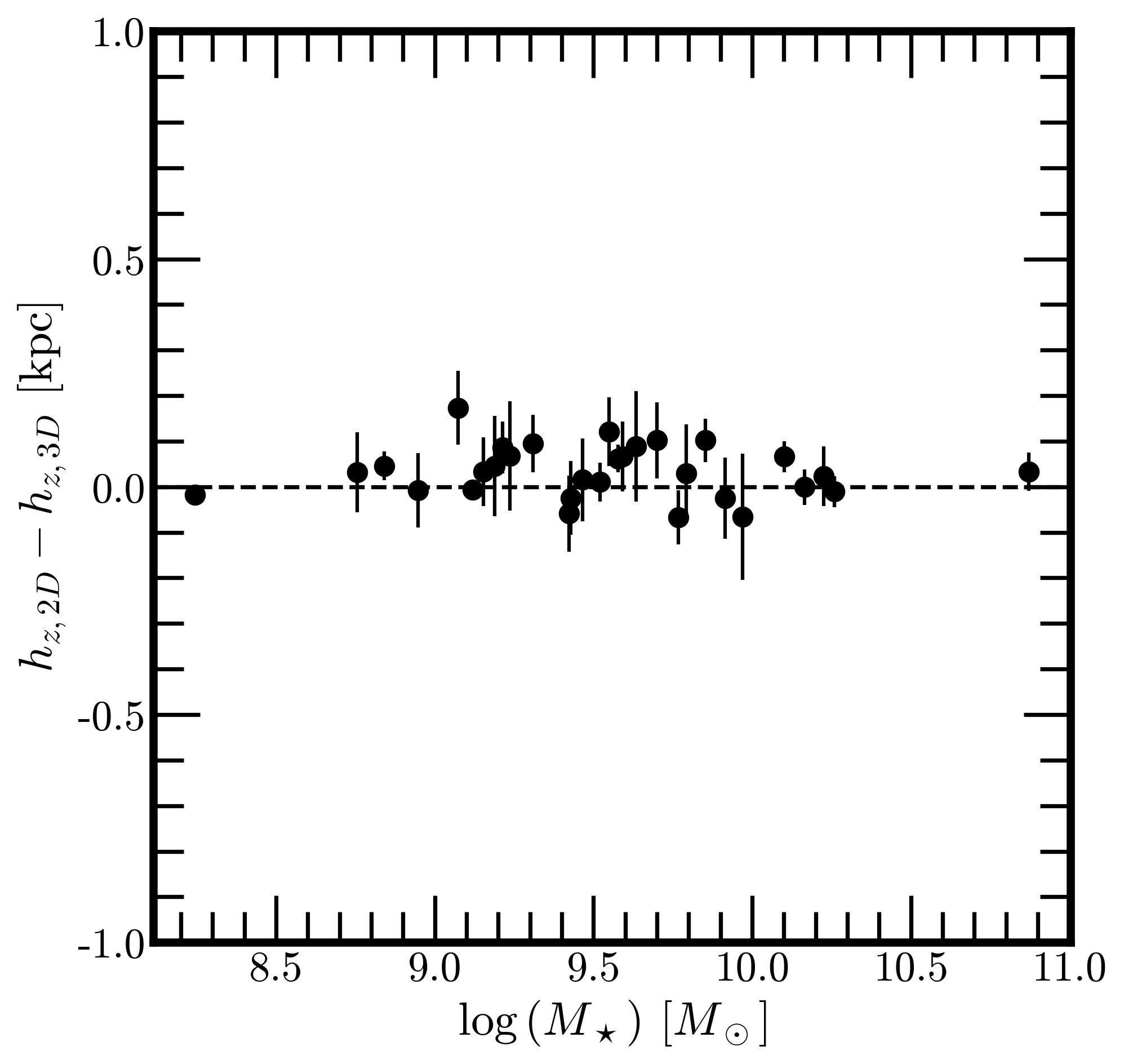}
	\caption{Comparison of scale height results from 3D and 2D methods.}
	\label{fig:2Dvs3Dresults}
\end{figure}

Figure \ref{fig:2Dvs3Dresults} shows a comparison between scale heights from the 2D and 3D methods. This shows that the two methods are in good agreement, displaying a smaller scatter around the identity line, compared to the plots in Figure \ref{fig:1Dvs2D} for the 1D and 2D methods . This indicates that the GALFIT and IMFIT models are consistent despite the added flexibility in profile shape that \textsc{Imfit} allows with the insertion of the parameter $ n $.

%\begin{figure}
%%	\vspace{-.25cm}
%	\centering
%	\includegraphics[width=\columnwidth]{Exp3D_hz_vs_corrected_galfit_2D_best_masked.png}
%	\caption{Comparison of scale height results from 3D and 2D methods.}
%	\label{fig:2Dvs3Dresults}
%\end{figure}
%\hfill

%\begin{figure*}
%%	\hspace{-1cm}
%	\centering
%	\includegraphics[width=1\textwidth]{2D_results/NGC4244_dat_mod_res.png}
%	\caption[Data, model and residual images of the NGC 4244 galaxy]{Data (top panel), model (middle panel) and residual (bottom panel) images of the NGC 4244 galaxy from 2D fits using \textsc{Galfit} EdgeOnDisk model. Images are in unit of counts and are scaled linearly with a Min Max type limit.}
%	\label{fig:NGC4244datmodres}
%\end{figure*}
%\makeatother
\begin{figure*}
	%	\hspace{-1cm}
	\centering
	\includegraphics[width=1\textwidth]{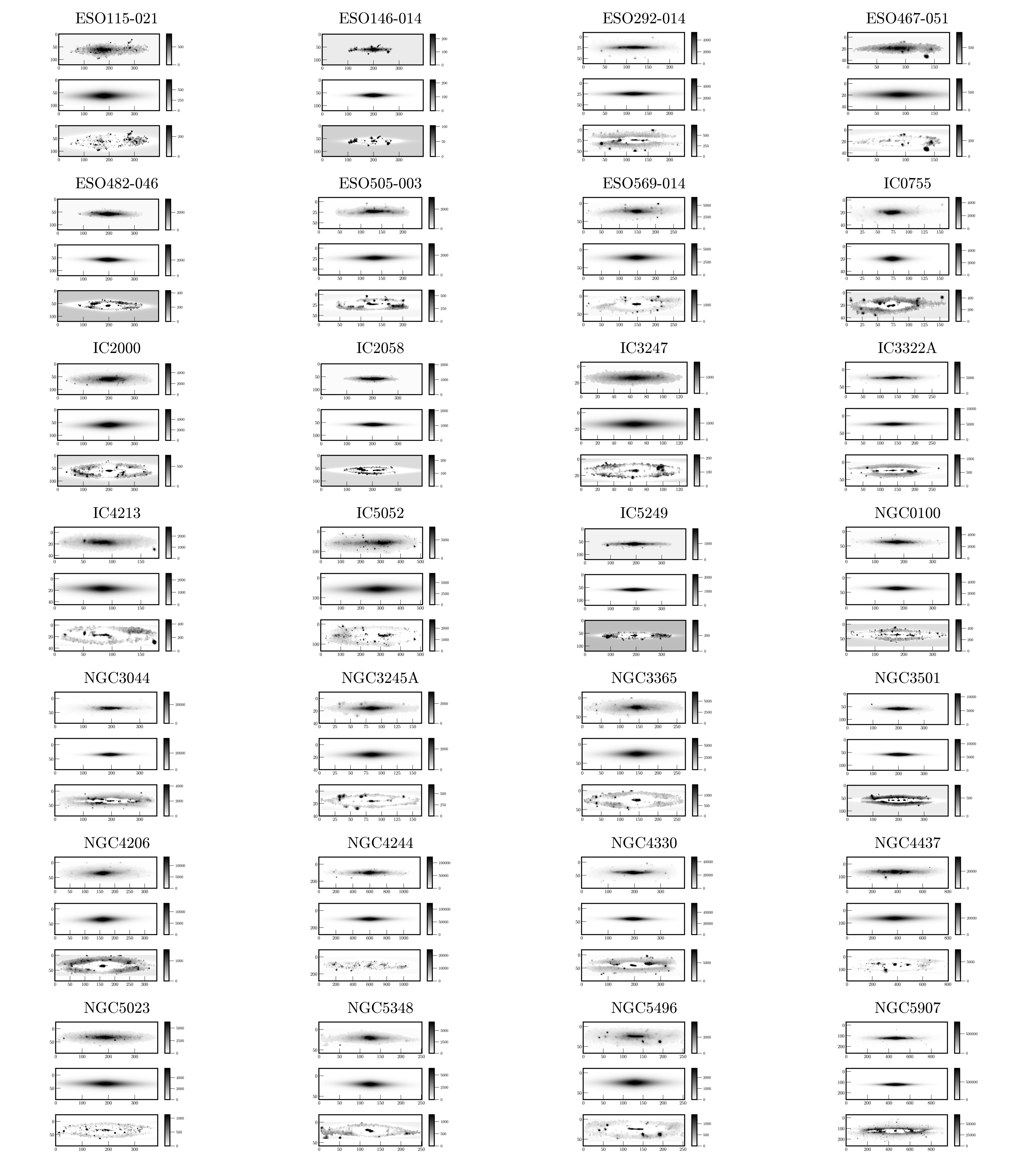}
	\caption{Data, model and residual images of our sample from 2D fits using \textsc{Galfit} EdgeOnDisk model. Images are in unit of counts with linear color distributions.}
	\label{fig:NGC4244datmodres}
\end{figure*}

\begin{figure*}
	%	\hspace{-1cm}
	\centering
  %  \ContinuedFloat
	\includegraphics[width=1\textwidth]{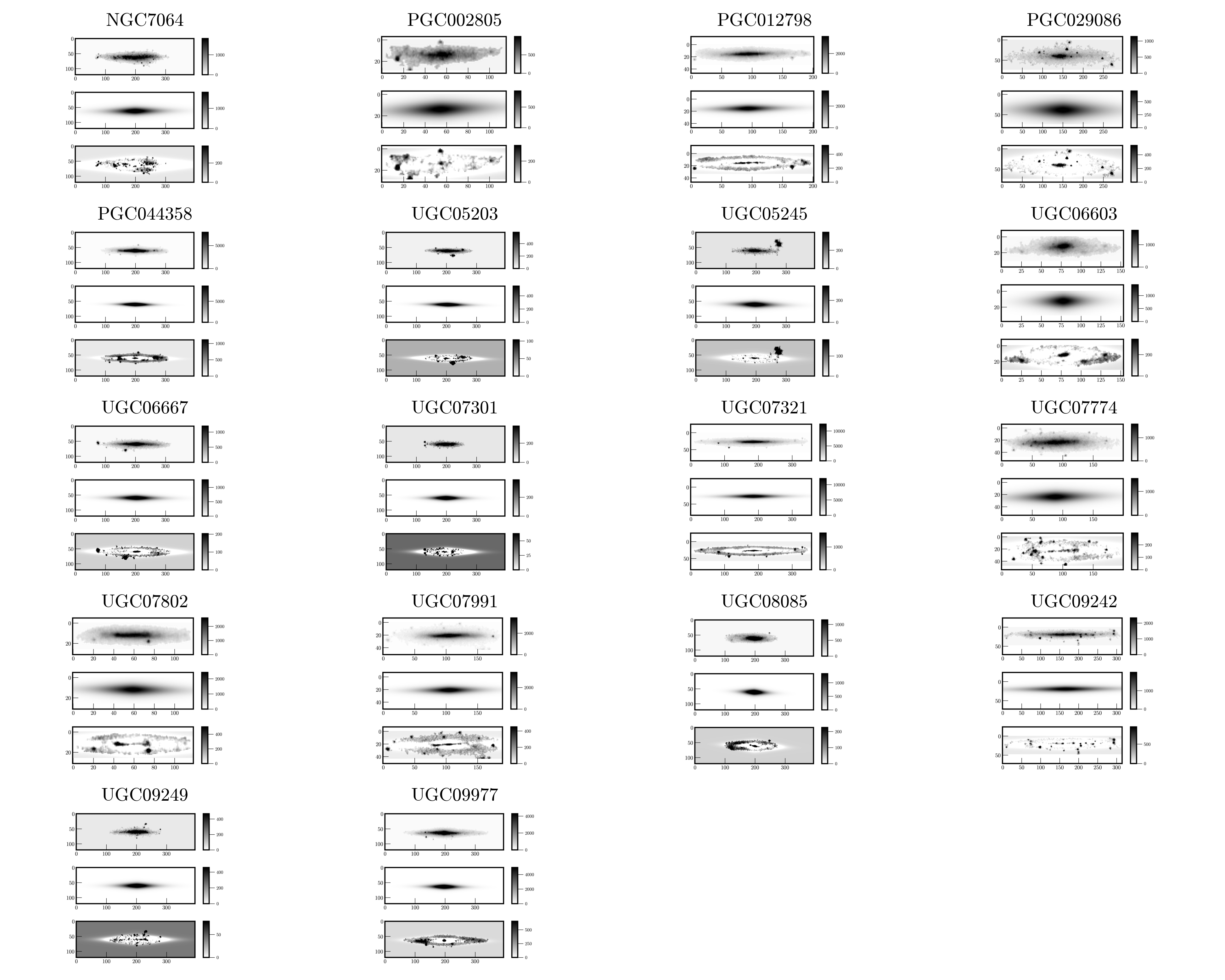}
	\caption{Cont.}
	\label{fig:NGC4244datmodres2}
\end{figure*}

\begin{table*}
	\small
	\centering
	\caption{Best-fit results from \textsc{Galfit} EdgeOnDisk model. \textit{Column 1}: Galaxy name. \textit{Column 2}: Inclination-corrected central surface brightness as observed in the 3.6 $ \mu  $m band. \textit{Column 3}: Uncertainty on the central surface brightness. \textit{Column 4}: Radial scale length in kiloparsecs. \textit{Column 5}:  Uncertainty on radial scale length. \textit{Column 6}: Inclination-corrected vertical scale height in kiloparsecs. \textit{Column 7}: Uncertainty on scale height.} 
	\label{tab:2Dresults}
	%\endhead
	\begin{tabular}{lcccccc}
		\hline
		\hline
		Galaxy & $ \mu (0,0)_{3.6\mu m} $ & $ \pm $ & $ h_R $ & $ \pm $ & $ h_z $ & $ \pm $  \\
		& [mag arcsec$^{-2}$] &  & [kpc] &  & [kpc] &  \\
		(1) & (2) & (3) & (4) & (5) & (6) & (7) \\
		\hline
		ESO 115-021&20.69&0.002&0.90&0.01&0.22&0.01\\
		ESO 146-014&20.89&0.003&1.87&0.17&0.26&0.02\\
		ESO 292-014&18.24&0.002&2.00&0.36&0.18&0.03\\
		ESO 467-051&20.51&0.008&2.39&0.48&0.33&0.07\\
		ESO 482-046&19.11&0.001&2.42&0.14&0.39&0.02\\
		ESO 505-003&18.73&0.003&1.93&0.18&0.33&0.03\\
		ESO 569-014&18.93&0.004&2.76&0.25&0.42&0.04\\
		IC 0755&17.78&0.003&0.98&0.24&0.20&0.05\\
		IC 2000&18.91&0.001&2.91&0.27&0.41&0.04\\
		IC 2058&18.95&0.001&1.98&0.35&0.23&0.04\\
		IC 3247&19.37&0.003&1.76&0.16&0.19&0.02\\
		IC 3322A&17.66&0.002&2.81&0.26&0.29&0.03\\
		IC 4213&19.03&0.004&1.77&0.37&0.23&0.05\\
		IC 5052&18.98&0.004&1.51&0.07&0.17&0.01\\
		IC 5249&20.22&0.002&3.44&0.32&0.61&0.06\\
		NGC 0100&18.84&0.001&2.16&0.20&0.25&0.02\\
		NGC 3044&16.91&0.002&1.70&0.16&0.25&0.02\\
		NGC 3245A&18.93&0.003&1.68&0.35&0.44&0.09\\
		NGC 3365&18.77&0.003&1.95&0.40&0.29&0.06\\
		NGC 3501&17.31&0.001&2.28&0.21&0.30&0.03\\
		NGC 4206&18.08&0.002&2.15&0.45&0.44&0.09\\
		NGC 4244&19.04&0.001&1.60&0.05&0.21&0.01\\
		NGC 4330&18.08&0.002&2.02&0.42&0.29&0.06\\
		NGC 4437&18.32&0.007&3.20&0.31&0.49&0.05\\
		NGC 5023&19.20&0.002&1.29&0.02&0.14&0.01\\
		NGC 5348&18.77&0.003&1.37&0.28&0.32&0.07\\
		NGC 5496&18.82&0.003&2.47&0.51&0.33&0.07\\
		NGC 5907&16.76&0.001&3.80&0.18&0.66&0.03\\
		NGC 7064&19.74&0.002&1.76&0.35&0.29&0.06\\
		PGC 002805&20.18&0.006&1.46&0.29&0.18&0.04\\
		PGC 012798&19.04&0.002&2.52&0.50&0.47&0.09\\
		PGC 029086&20.75&0.007&2.82&0.56&0.44&0.09\\
		PGC 044358&17.70&0.001&1.67&0.33&0.20&0.04\\
		UGC 05203&20.19&0.002&2.93&0.58&0.35&0.07\\
		UGC 05245&21.07&0.004&3.12&0.62&0.76&0.15\\
		UGC 06603&19.52&0.004&1.51&0.30&0.29&0.06\\
		UGC 06667&19.56&0.001&2.14&0.42&0.27&0.05\\
		UGC 07301&20.38&0.003&2.22&0.44&0.25&0.05\\
		UGC 07321&19.54&0.002&3.04&0.60&0.41&0.08\\
% It seems like I need to have 
		UGC 07774&19.82&0.002&3.05&0.60&0.36&0.07\\
		UGC 07802&19.03&0.004&1.39&0.27&0.15&0.03\\
		UGC 07991&18.60&0.002&1.6&0.32&0.16&0.03\\
		UGC 08085&18.85&0.002&2.01&0.4&0.32&0.06\\
	
		UGC 09242&20.84&0.005&5.50&1.09&0.62&0.12\\
		UGC 09249&20.40&0.003&2.11&0.52&0.26&0.06\\
		UGC 09977&18.64&0.002&2.32&0.46&0.40&0.08\\
		\hline
		\hline
	\end{tabular}
	
\end{table*}
\subsection{Thin and thick disk components}\label{sec:res2comp}

% due to the bright infrared sky, the near-infrared images are of lower S/ N and cannot reach to faint regions where a thick disk would dominate.
%
%This would mean that thick disks might be less prominent for higher morphological type T. 
% It has been established that disk galaxies have a thin stellar disk with gas and dust (\citealt{}aside from the more thicker stellar disk component.

We use the 1D sech$^2$ model to investigate two-component disks. Best-fit parameters are shown in Table \ref{tab:1Dsech2results}. We find that close to two-thirds of our galaxies show the presence of a thick disk considering our sample is composed of nearby late-type spiral galaxies. In comparison, \cite{2018A&A...610A...5C} found that thick disks are always present using a sample of mainly distant and early-type galaxies. It has been established for the Milky Way that the thin disk has two distinct components: a ``young thin disk'' dominated by young OB associations and an ``old'' thin disk with older redder stars (\citetalias{2006AJ....131..226Y}). The young thin disk, the old thin disk and the thick disk of the Milky Way usually have scale heights of $\sim 0.1 \ \text{kpc} ,\ \sim 0.3 \ \text{kpc}$, and $ \sim 1 \ \text{kpc} $, respectively (\citealp{1993ApJ...409..635R}; \citealp{2003AJ....125.1958L}). However, all of the thick disks in our sample have $h_{z \ \text{thick}} < 1$ kpc and a few galaxies have  $h_{z \ \text{thin}} \sim 0.1$ kpc suggesting that the two-component disk we find could correspond to the young and old thin disks. Furthermore, our near-infrared insight at 3.6 $ \mu $m provides us with a is dust-free view and is likely to contain most of the old stellar population. This implies that the two components we found are indeed the old thin disk and the thick old stellar disk. Our average scale heights of $\bar{h}_{z \ \text{thin}} = 0.14 \pm 0.07$ kpc and $\bar{h}_{z \ \text{thick}} = 0.33 \pm 0.16$ kpc are smaller than the dust-free measurements reported by \citetalias{2006AJ....131..226Y} ($\bar{h}_{z \ \text{thin}} \sim 0.27$ kpc; $\bar{h}_{z \ \text{thick}} \sim 0.59$ kpc) using sech$ ^2 $ models on B-, R- and K-band data. However, the fact that our inclination-uncorrected scale heights ($\bar{h}_{z \ \text{thin}} = 0.22 \pm 0.07$ kpc; $\bar{h}_{z \ \text{thick}} = 0.52 \pm 0.18$ kpc) matched those from \citetalias{2006AJ....131..226Y} which does not explicitly mention an inclination correction implies that the inconsistency might be due to projection effect. We also compare our scale heights with those from \citet{2011ApJ...729...18C, 2011ApJ...741...28C}, who examined the vertical structure of nearby galaxies using S$^4$G images, with a few galaxies overlapping with our sample. For example, our two-component fits for NGC 4244 were unsuccessful, whereas \citet{2011ApJ...729...18C} identified a faint thick disk on the galaxy's near side using a single-component exponential model across varying heights $z$. While we find a single-component $h_z = 0.21$ kpc (uncorrected for inclination of $h_z = 0.35$ kpc), they measured an average $h_{z \ \text{thin}} \sim 0.35$ kpc for this galaxy. For the other overlapping galaxies, such as NGC 3501 the measured scale height of ($h_{z \ \text{thin}} = 0.15$ kpc; $h_{z \ \text{thick}} = 0.36$ kpc inclination corrected) and (uncorrected for inclination, $h_{z \ \text{thin}} = 0.22$ kpc; $h_{z \ \text{thick}} = 0.51$ kpc) are in agreement with results from \citet{2011ApJ...741...28C} (average $h_{z \ \text{thin}} \sim 0.20$ kpc; $h_{z \ \text{thick}} \sim 0.59$ kpc). On the other hand, the scale height of NGC 4330 is lower compared to the literature (e.g \citealt{2011ApJ...729...18C}). We measured a scale height of $h_{z \ \text{thin}} = 0.11$ kpc; $h_{z \ \text{thick}} = 0.31$ kpc (uncorrected for inclination of $h_{z \ \text{thin}} = 0.16$ kpc; $h_{z \ \text{thick}} = 0.45$ kpc) for this galaxy compared to their measurements (average $h_{z \ \text{thin}} \sim 0.18$ kpc; $h_{z \ \text{thick}} \sim 0.74$ kpc) from  \citet{2011ApJ...741...28C}. This discrepancy is likely due to the difference in inclination correction and model fitting as  \citet{2011ApJ...741...28C} apply separate corrections for each disk component and utilize a hydrostatic equilibrium solution model instead of the simpler sech$^2$ function.

\begin{figure}
	%	\vspace{-0.5cm}
	\centering
	\includegraphics[width=1\columnwidth]{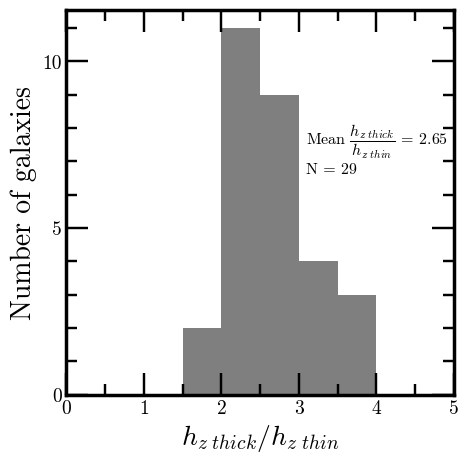}
	\caption{Distribution of $h_z$ ratio.}
	\label{fig:hist_hz_ratio}
\end{figure}

\begin{figure}
	%	\vspace{-0.5cm}
	\centering
	\includegraphics[width=1\columnwidth]{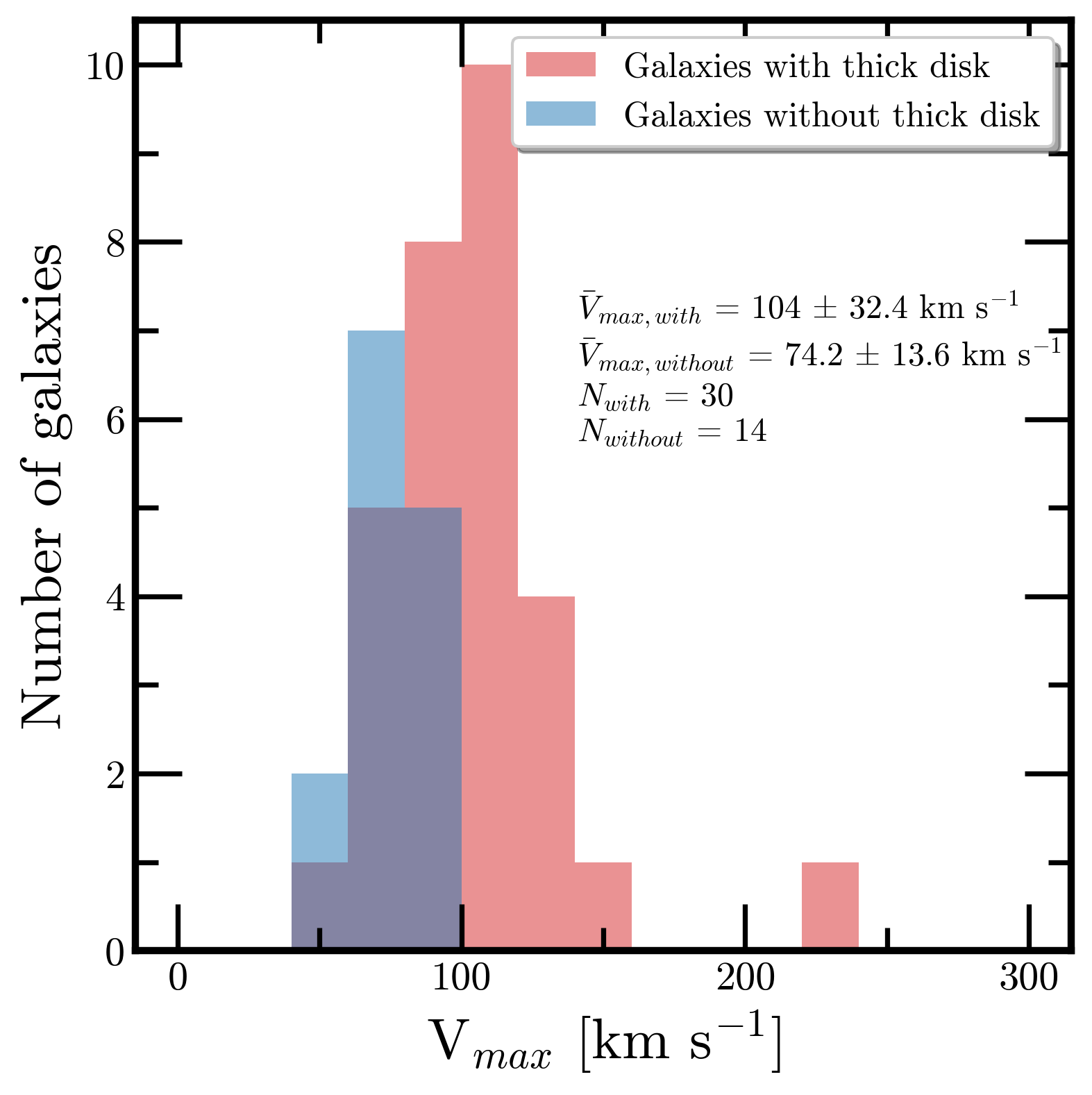}
	\caption{Histogram of $V_\text{max}$ for galaxies showing the presence of a thick disk (blue) and those that do not possess such a component (red).}
	\label{fig:hist_Vmax}
\end{figure}

Ratios of thick and thin disk scale heights $h_{z \ \text{thick}}/h_{z \ \text{thin}}$ are also presented in Table \ref{tab:1Dsech2results}. We find an average $h_{z \ \text{thick}}/h_{z \ \text{thin}} = 2.65 \pm 0.56$. In comparison, \citetalias{2006AJ....131..226Y} find an average $h_{z \ \text{thick}}/h_{z \ \text{thin}} = 2.5 $. It is important to reiterate that we choose to use the near-infrared data in this paper in order to better trace the older stellar populations that dominate these stellar structures. The distribution of scale height ratios is shown in Figure \ref{fig:hist_hz_ratio}. Because the ratio of thick-to-thin disk stellar mass has been shown to be related to the global kinematics of galaxies (\citealt{2022ApJ...932...28V}), it is also interesting to probe for a correlation between $h_{z \ \text{thick}}/h_{z \ \text{thin}}$ and $V_\text{flat}$ or the maximum velocity $ V_{\max} $. Figure \ref{fig:hist_Vmax} illustrates that galaxies with a thick disk component have higher $V_\text{max}$ values (on average $104 \pm 32.4 \text{km s}^{-1}$) than those without this disk feature (on average $74.2 \pm 13.6 \text{km s}^{-1}$). This observation confirms the correlation between dynamical mass and disk accretion. This aligns with the evidence suggesting the formation of massive galaxies through mergers, as noted in \citet{2011A&A...530A..10Q}, which implies the presence of a thick disk.
%The reader is referred to Section \ref{sec:correlations} for the correlations with other galaxy properties are presented.   

Figure \ref{fig:UGC09977_vertical_profile} shows the successful vertical disk profile fits using two sech$^2$ models. It can be observed that both disks are generally approximately symmetrical relative to the galactic mid-plane as most galaxies of our sample. The averaged profile is represented by the squared data points, the red solid line is the total best-fit model, the dashed blue line corresponds to the thin disk model and the dotted and dashed green line is the thick disk model.
%Fits were performed within an apparent vertical range which causes 
%Fits returned a reduced chi-squared value $\chi_\nu^2 = 0.39 $ which  

\begin{figure*}
	%	\vspace{-0.5cm}
	\centering
	\includegraphics[width=1\textwidth]{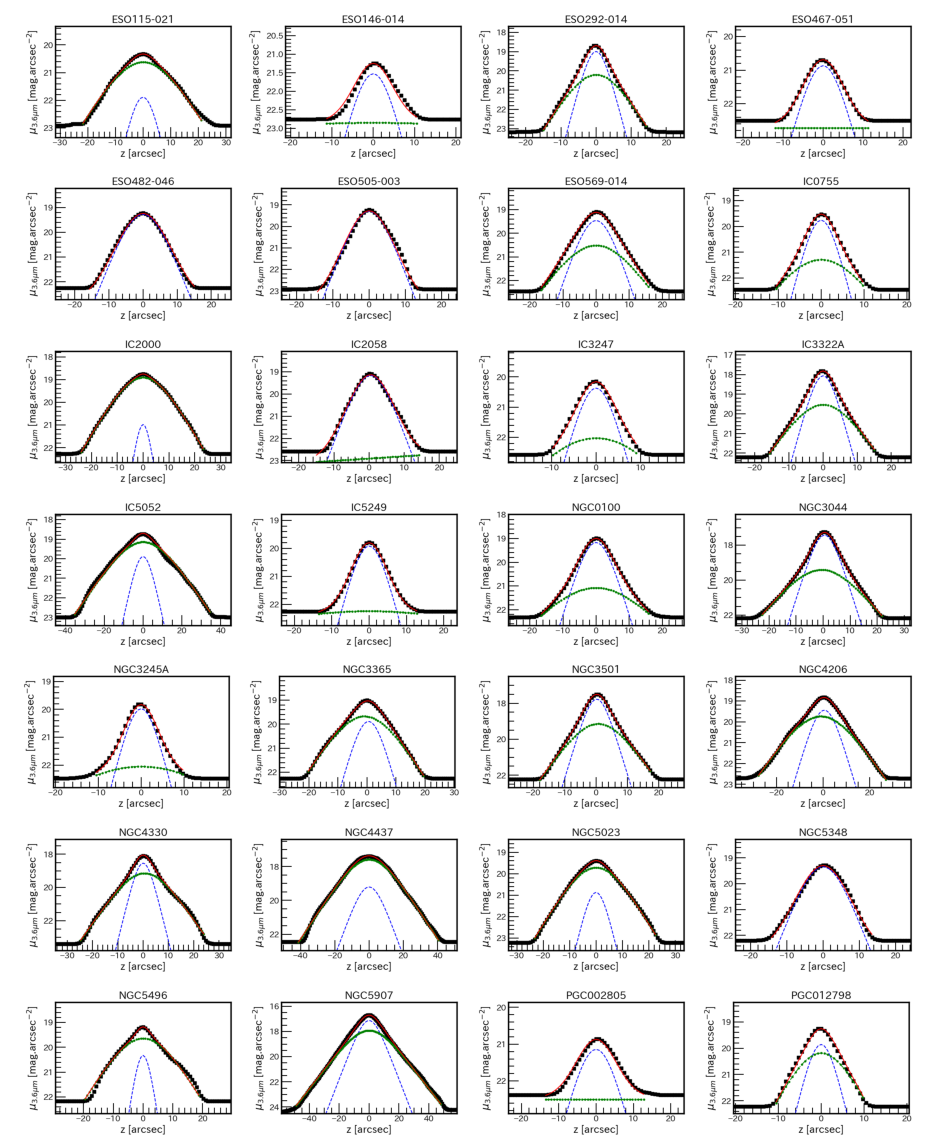}
	\caption{Successful vertical fits of our sample. Squared points represent 3.6 $ \mu $m data. The red solid line is the total best-fit model composed of two distinct components: the thin disk component (dashed blue line) and the thick disk component (dotted and dashed green line). See Section \ref{sec:res2comp} for details.}
	\label{fig:UGC09977_vertical_profile}
\end{figure*}
\begin{figure*}
	%	\vspace{-0.5cm}
  %  \ContinuedFloat
	\centering
	\includegraphics[width=1\textwidth]{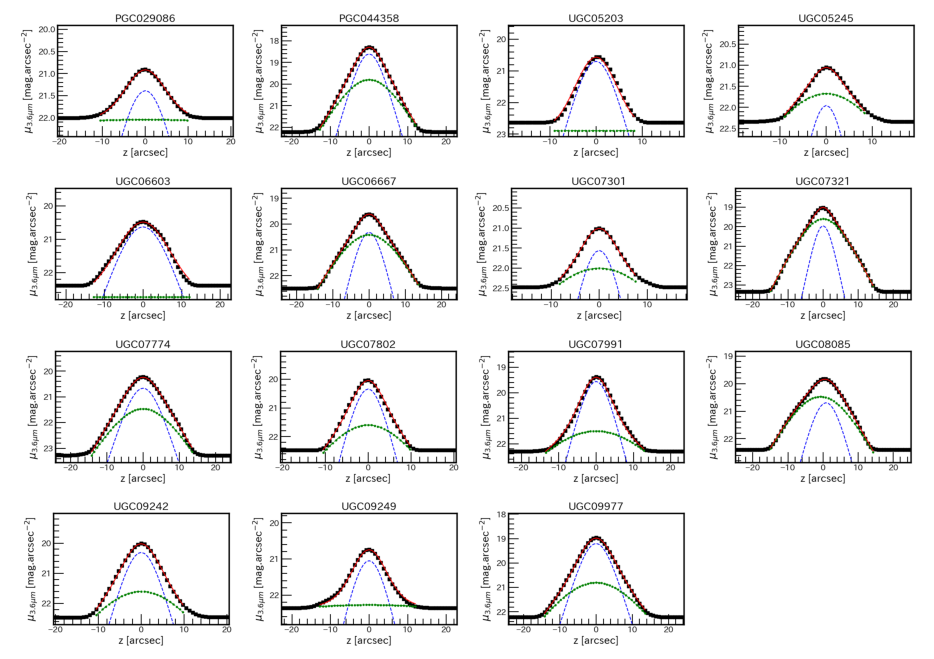}
	\caption{Cont.}
	\label{fig:UGC09977_vertical_profile2}
\end{figure*}
\begin{table*}
	\small
	%\centering
	%	\hspace{-3cm}
	\caption{Best-fit results from one-dimensional two-component sech$^2$ model. \textit{Column 1}: Galaxy name. \textit{Column 2}: Thin disk average central surface brightness as observed in the 3.6 $ \mu  $m band. \textit{Column 3}: Uncertainty on thin disk average central surface brightness. \textit{Column 4}: Inclination-corrected thin disk vertical scale height. \textit{Column 5}: Uncertainty on thin disk scale height. \textit{Column 6}: Thick disk average central surface brightness. \textit{Column 7}: Uncertainty on thick disk average central surface brightness. \textit{Column 8}: Inclination-corrected thick disk vertical scale height. \textit{Column 9}: Uncertainty on thick disk scale height. \textit{Column 10}: Ratio of thick and thin disk scale height.} 
	\label{tab:1Dsech2results}
	%\endhead
	\begin{tabular}{lccccccccc}
		\hline
		\hline
		Galaxy & $ \mu (0)_\text{thin} $ & $ \pm $ & $ h_{z \ \text{thin}} $ & $ \pm $ & $ \mu (0)_\text{thick} $ & $ \pm $ & $ h_{z \ \text{thick}} $ & $ \pm $ & $ \frac{h_{z \ \text{thick}}}{h_{z \ \text{thin}}} $ \\
		& $\rm [mag \ arcsec^{2}]$ & & [kpc] &  & $\rm [mag \ arcsec^{2}]$ & & [kpc] &  & \\
		(1) & (2) & (3) & (4) & (5) & (6) & (7) & (8) & (9) & (10)\\
		\hline
		ESO 115-021&21.9&0.21&0.05&0.01&20.6&0.07&0.13&0.01&2.69\\
		ESO 146-014&21.5&0.16&0.13&0.02&--&--&--&--&--\\
		ESO 292-014&19.0&0.08&0.09&0.02&20.2&0.23&0.22&0.04&2.37\\
		ESO 467-051&20.9&0.01&0.18&0.04&--&--&--&--&--\\
		ESO 482-046&19.3&0.02&0.21&0.01&--&--&--&--&--\\
		ESO 505-003&19.3&0.05&0.19&0.02&--&--&--&--&--\\
		ESO 569-014&19.5&0.13&0.18&0.02&20.5&0.39&0.38&0.06&2.06\\
		IC 0755&19.8&0.05&0.12&0.03&21.3&0.24&0.34&0.09&2.79\\
		IC 2000&21.0&0.10&0.06&0.01&18.9&0.01&0.24&0.02&3.85\\
		IC 2058&19.2&0.03&0.13&0.02&--&--&--&--&--\\
		IC 3247&20.4&0.16&0.11&0.02&--&--&--&--&--\\
		IC 3322A&18.1&0.05&0.14&0.02&19.5&0.17&0.36&0.04&2.51\\
		IC 5052&19.9&0.13&0.03&0.01&19.1&0.04&0.09&0.01&3.11\\
		IC 5249&19.9&0.01&0.33&0.03&--&--&--&--&--\\
		NGC 0100&19.2&0.02&0.11&0.01&21.1&0.12&0.35&0.04&3.03\\
		NGC 3044&17.4&0.02&0.15&0.01&19.4&0.07&0.39&0.04&2.65\\
		NGC 3245A&20.0&0.05&0.23&0.05&--&--&--&--&--\\
		NGC 3365&19.9&0.07&0.11&0.02&19.7&0.06&0.24&0.05&2.26\\
		NGC 3501&17.8&0.02&0.15&0.01&19.1&0.08&0.36&0.03&2.34\\
		NGC 4206&19.4&0.04&0.17&0.04&19.7&0.05&0.41&0.09&2.37\\
		NGC 4330&18.5&0.06&0.11&0.02&19.1&0.06&0.31&0.06&2.82\\
		NGC 4437&19.2&0.51&0.13&0.04&17.6&0.13&0.23&0.02&1.80\\
		NGC 5023&20.9&0.12&0.03&0.01&19.7&0.04&0.08&0.01&2.55\\
		NGC 5348&19.3&0.04&0.18&0.04&--&--&--&--&--\\
		NGC 5496&20.3&0.20&0.07&0.02&19.6&0.06&0.26&0.05&3.94\\
		NGC 5907&17.1&0.05&0.29&0.02&17.9&0.09&0.58&0.03&2.01\\
		PGC 002805&21.1&0.04&0.11&0.02&--&--&--&--&--\\
		PGC 012798&19.9&0.36&0.21&0.06&20.2&0.53&0.45&0.11&2.15\\
		PGC 029086&21.4&0.10&0.09&0.02&--&--&--&--&--\\
		PGC 044358&18.6&0.08&0.11&0.02&19.8&0.28&0.23&0.05&2.09\\
		UGC 05203&20.7&0.09&0.19&0.04&--&--&--&--&--\\
		UGC 05245&22.0&0.21&0.24&0.06&21.7&0.17&0.78&0.2&3.26\\
		UGC 06603&20.6&0.07&0.17&0.03&--&--&--&--&--\\
		UGC 06667&20.3&0.05&0.10&0.02&20.4&0.06&0.23&0.05&2.40\\
		UGC 07301&21.6&0.07&0.11&0.02&22.0&0.12&0.37&0.09&3.48\\
		UGC 07321&20.0&0.08&0.14&0.03&19.6&0.05&0.35&0.07&2.45\\
		UGC 07774&20.7&0.14&0.15&0.03&21.5&0.31&0.30&0.07&1.96\\
		UGC 07802&20.3&0.10&0.08&0.02&21.6&0.34&0.23&0.06&2.79\\
		UGC 07991&19.6&0.03&0.09&0.02&21.5&0.14&0.32&0.07&3.76\\
		UGC 08085&20.7&0.10&0.14&0.03&20.5&0.09&0.29&0.06&2.13\\
		UGC 09242&20.3&0.10&0.26&0.05&21.6&0.36&0.72&0.21&2.73\\
		UGC 09249&21.0&0.05&0.11&0.03&--&--&--&--&--\\
		UGC 09977&19.2&0.03&0.20&0.04&20.8&0.16&0.51&0.11&2.52\\
		%		\hline
		%	\end{tabular}
		
		%\end{table*}
		%\newpage
		%\vspace{.5in}
		%\makeatletter
		%\setlength{\@fptop}{0pt}
		%%	\pagebreak
		%\begin{table*}[t!]
		%	\centering
		%	\small
		%	\begin{tabular}{lccccccccc}
		%		\hline
		%		Galaxy & $ \mu (0)_\text{thin} $ & $ \pm $ & $ h_{z \ \text{thin}} $ & $ \pm $ & $ \mu (0)_\text{thick} $ & $ \pm $ & $ h_{z \ \text{thick}} $ & $ \pm $ & $ \frac{h_{z \ \text{thick}}}{h_{z \ \text{thin}}} $ \\
		%		& $\left[\frac{\text{mag}}{\text{arcsec}^{2}}\right]$ & & [kpc] &  & $\left[\frac{\text{mag}}{\text{arcsec}^{2}}\right]$ & & [kpc] &  & \\
		%		(1) & (2) & (3) & (4) & (5) & (6) & (7) & (8) & (9) & (10)\\
		%		\hline
		%		UGC 07774&20.7&0.14&0.15&0.03&21.5&0.31&0.30&0.07&1.96\\
		%		UGC 07802&20.3&0.10&0.08&0.02&21.6&0.34&0.23&0.06&2.79\\
		%		UGC 07991&19.6&0.03&0.09&0.02&21.5&0.14&0.32&0.07&3.76\\
		%		UGC 08085&20.6&0.42&0.14&0.04&20.6&0.42&0.30&0.07&2.05\\
		%		UGC 09242&20.3&0.10&0.26&0.05&21.6&0.36&0.72&0.21&2.73\\
		%		UGC 09249&21.0&0.05&0.11&0.03&$-$&$-$&$-$&$-$&$-$\\
		%		UGC 09977&19.2&0.03&0.2&0.04&20.8&0.16&0.51&0.11&2.52 \\
		\hline
		\hline
	\end{tabular}	
\end{table*}

\subsection{The flaring of disks}\label{sec:diskflaring}
The radial profiles of the vertical disk scale heights are shown in Figure \ref{fig:UGC09977radial_profile_of_hz_sech} to \ref{fig:radial_profile_of_hz_sech_part3} and  the profiles of all the galaxies in our sample are presented in Figure \ref{fig:radial_profile_of_hz_all}. These figures show that almost all the galaxies in our sample have an increasing disk thickness from the center to the outer part of the galaxies. The flaring amplitude given as ($h_{z, R_{25}}-h_{z, R=0})/h_{z, R=0}$ varies for each galaxy but only two galaxies (IC5052 and NGC5023) have less than 10 \% disk flaring amplitude. Therefore, outer disk flaring is observed for all the galaxies in our sample. However, using observation of edge-on galaxies in B, V, R and I band imaging, \citealt{1996A&AS..117...19D} noted that the observed disk flaring could be attributed to optical edge effect and warps. Furthermore, \citet{1996A&AS..117...19D} found through visual inspection of the radial profile of the scale heights that the flaring is more important in terms of amplitude in early-type disks than in later-type ones. To check if our sample follows this tendency, we plot the ratio of the scale height at the apparent radius and the galaxy center $h_{z, R_{25}}/h_{z, R=0}$ against the morphological type T (Figure \ref{fig:r25_r0_vs_T}). Note that the apparent radius $R_{25}$ (indicated in Figure \ref{fig:UGC09977radial_profile_of_hz_sech}) has been measured in the B-band and retrieved from the Hypercat-Lyon-Meudon Extragalactic Database\footnote{\url{http://leda.univ-lyon1.fr}} (HyperLEDA, \citealp{2014A&A...570A..13M}). As proved by \cite{1996A&AS..118..557D}, galaxies tend to be bluer with increasing radius which means that $R_{25}$ measured in the B-band might be bigger than the apparent radius as seen in the 3.6 $ \mu\text{m}$ imaging. It can be seen from Figure \ref{fig:r25_r0_vs_T} that the increase in scale height is indeed more evident in earlier type galaxies than in later type ones (with correlation coefficients $R_{Pearson}=-0.32 $ and $R_{Spearman}=-0.29 $, a p-value of 0.03 and given the fact that only 4 galaxies of our sample have $T < 5$). 
The best-fit relation is:
%$ h_z (R_{25})/h_z (R=0) = -0.13 \ T + 2.37 \pm 0.36 $.
\begin{equation}
	h_z (R_{25})/h_z (R=0) = -0.13 \ \text{T} + 2.37 \pm 0.36
\end{equation} 
with $h_z (R_{25})/h_z (R=0)$ the disk flaring amplitude and T the morphological type. This trend is thought to be caused by the change in bulge contribution to the total light of the galaxy \citet{1996A&AS..117...19D} as it evolves along the Hubble sequence. However, most of our galaxies do not have a prominent bulge and the bulge contamination usually affects the inner part of the disk therefore should not influence the flaring of the outer disk. In fact, observations from \citet{2016MNRAS.462.1470L} show that disks systematically exceed bulges in terms of size by a factor of $\sim 0.12$ to 0.69 dex for spiral galaxies considering all stellar masses between $\sim 10^7 - 10^{12} \ M_\odot$. Nonetheless, the effect of bulge contamination requires a more in-depth investigation with a more exhaustive sample.

\begin{figure*}
	%\vspace{-0.5cm}
	\centering
	\includegraphics[width=1\linewidth]{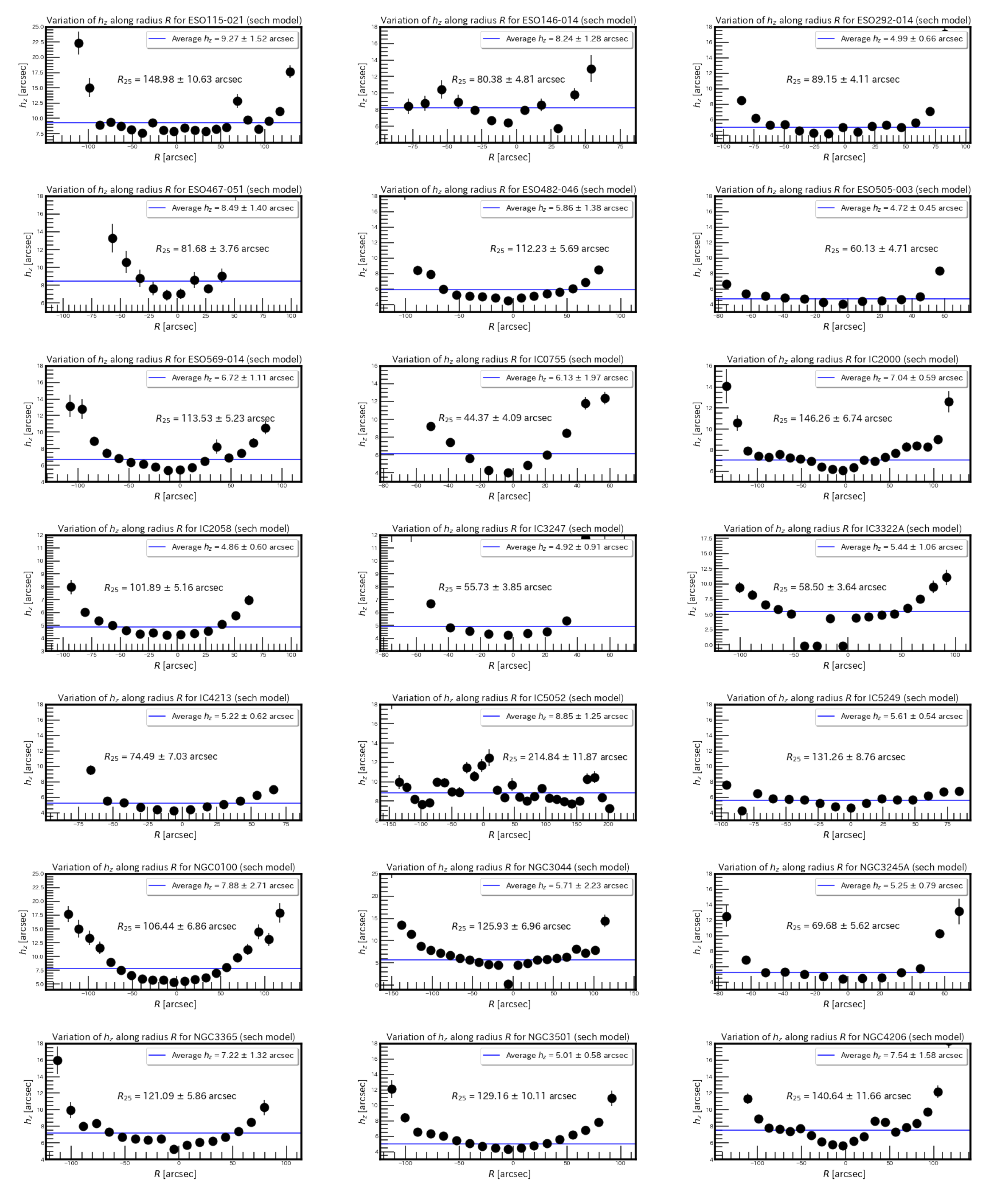}
	\caption{Radial profile of the scale height for our sample. Error bars represent statistical uncertainties from fits.}
%	\caption[Radial profile of the scale height for the UGC 09977 galaxy]{Radial profile of the scale height for the UGC 09977 galaxy. \textit{Top panel}: The applied box model of the galaxy. Each bin has a size of $8 \times 4$ pixels. \textit{Bottom panel}: The distribution of the scale height along the radius. Error bars represent statistical uncertainties from fits.}
	\label{fig:UGC09977radial_profile_of_hz_sech}
\end{figure*}
\begin{figure*}
	%\vspace{-0.5cm}
  %  \ContinuedFloat
	\centering
	\includegraphics[width=1\linewidth]{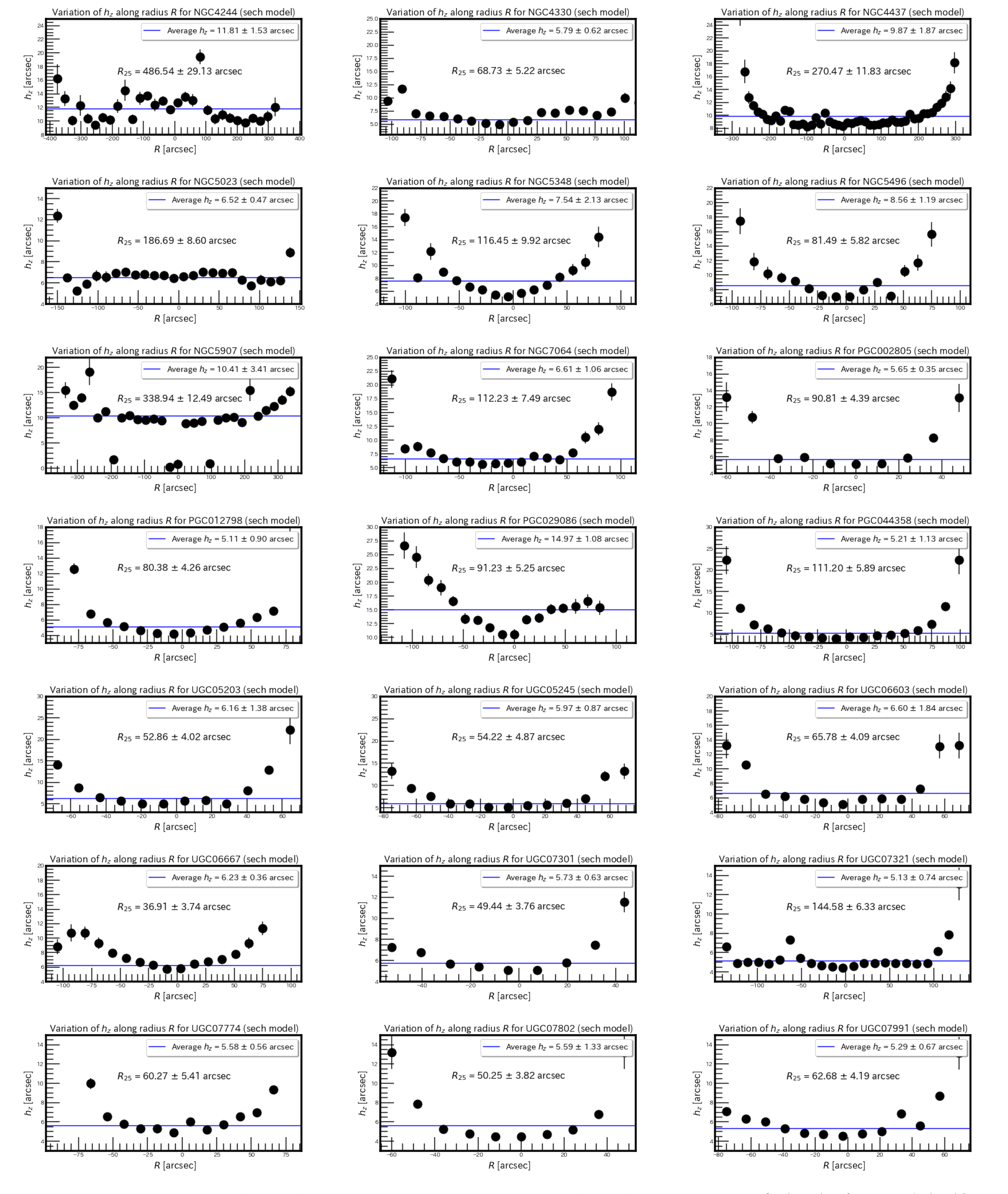}
	\caption{Cont.}
	%	\caption[Radial profile of the scale height for the UGC 09977 galaxy]{Radial profile of the scale height for the UGC 09977 galaxy. \textit{Top panel}: The applied box model of the galaxy. Each bin has a size of $8 \times 4$ pixels. \textit{Bottom panel}: The distribution of the scale height along the radius. Error bars represent statistical uncertainties from fits.}
	\label{fig:radial_profile_of_hz_sech_part2}
\end{figure*}

\begin{figure*}
	%\vspace{-0.5cm}
 %   \ContinuedFloat
	\centering
	\includegraphics[width=1\linewidth]{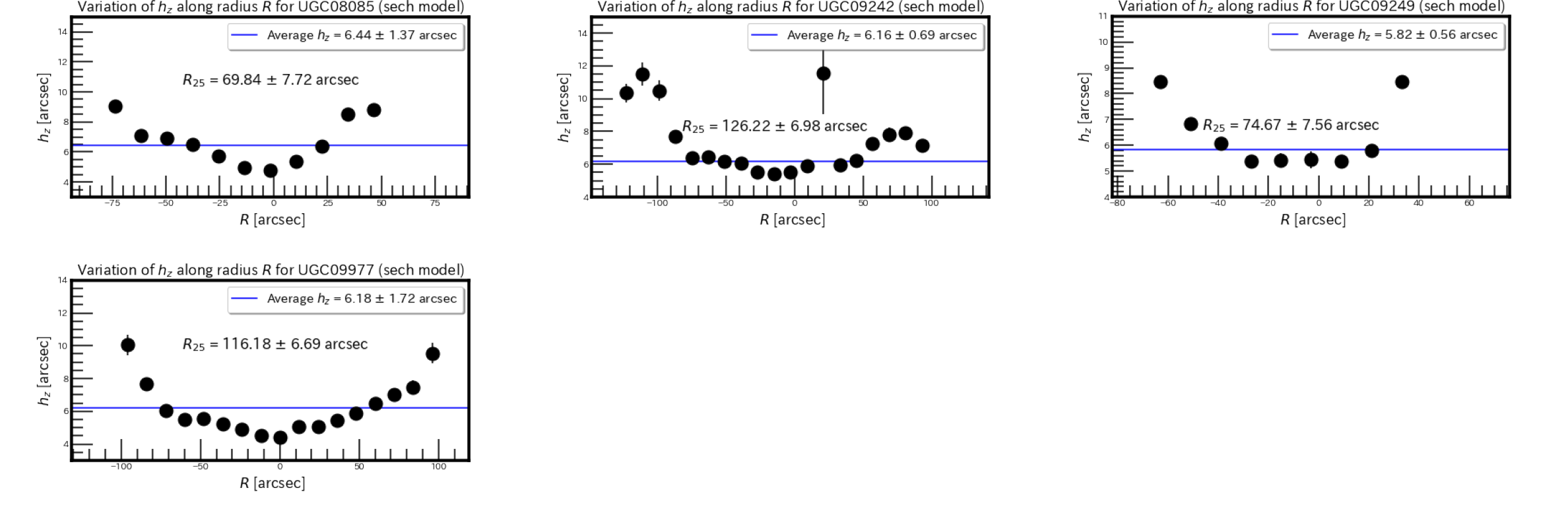}
	\caption{Cont.}
	%	\caption[Radial profile of the scale height for the UGC 09977 galaxy]{Radial profile of the scale height for the UGC 09977 galaxy. \textit{Top panel}: The applied box model of the galaxy. Each bin has a size of $8 \times 4$ pixels. \textit{Bottom panel}: The distribution of the scale height along the radius. Error bars represent statistical uncertainties from fits.}
	\label{fig:radial_profile_of_hz_sech_part3}
\end{figure*}

\begin{figure*}
	%\vspace{-0.5cm}
	\centering
	\includegraphics[width=1\linewidth]{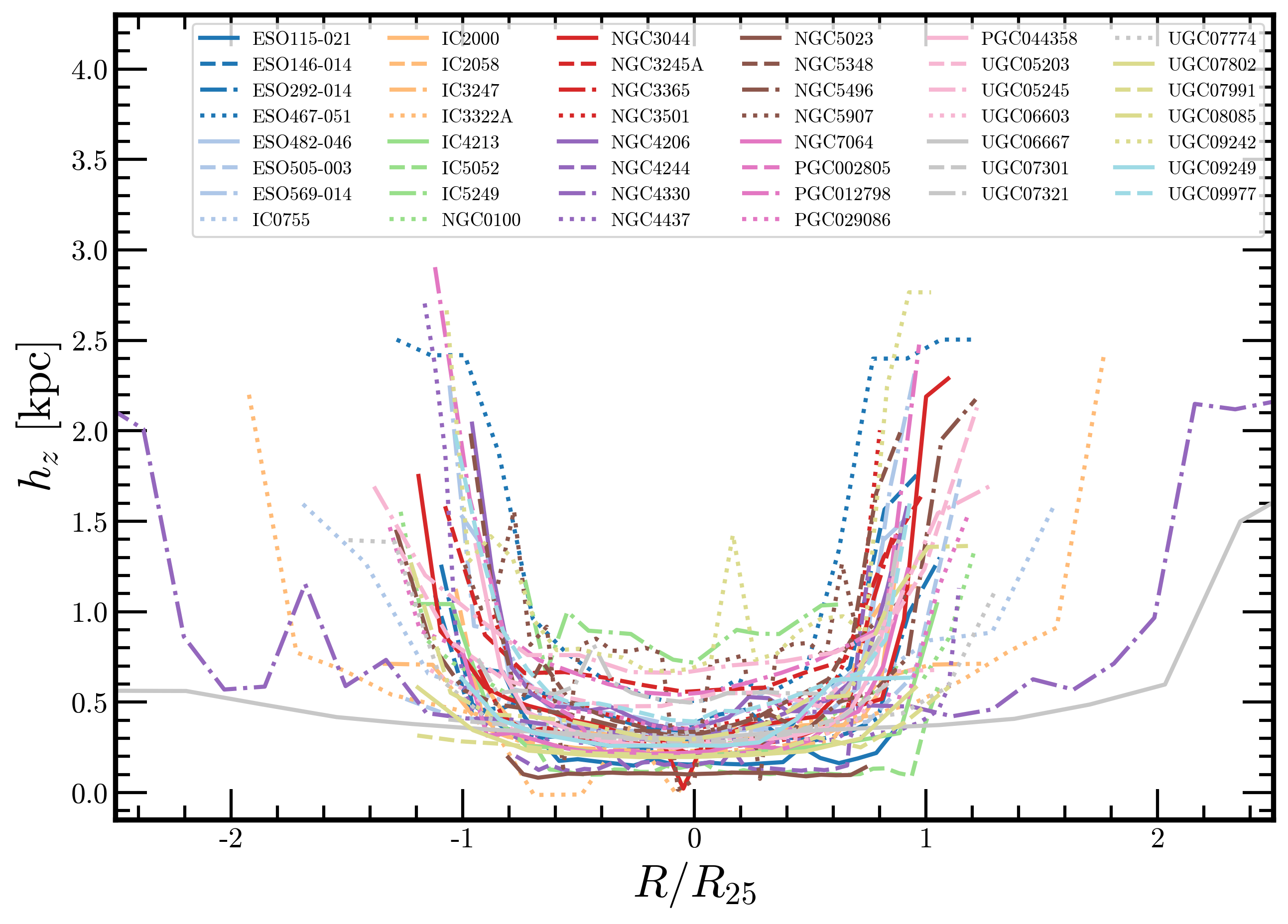}
	\caption{Radial profile of the scale height for our sample. Radii are normalized to the optical radius $R_{25}$.}
%	\caption[Radial profile of the scale height for the UGC 09977 galaxy]{Radial profile of the scale height for the UGC 09977 galaxy. \textit{Top panel}: The applied box model of the galaxy. Each bin has a size of $8 \times 4$ pixels. \textit{Bottom panel}: The distribution of the scale height along the radius. Error bars represent statistical uncertainties from fits.}
	\label{fig:radial_profile_of_hz_all}
\end{figure*}

% \begin{figure*}
% 	%\vspace{-0.5cm}
%   %  \ContinuedFloat
% 	\centering
% 	\includegraphics[width=1\linewidth]{radial_profiles_of_hz_atlas_part2.png}
% 	\caption{Cont.}
% 	%	\caption[Radial profile of the scale height for the UGC 09977 galaxy]{Radial profile of the scale height for the UGC 09977 galaxy. \textit{Top panel}: The applied box model of the galaxy. Each bin has a size of $8 \times 4$ pixels. \textit{Bottom panel}: The distribution of the scale height along the radius. Error bars represent statistical uncertainties from fits.}
% 	\label{fig:radial_profile_of_hz_sech_part2}
% \end{figure*}
% \begin{figure*}
% 	%\vspace{-0.5cm}
%  %   \ContinuedFloat
% 	\centering
% 	\includegraphics[width=1\linewidth]{radial_profiles_of_hz_atlas_part3.png}
% 	\caption{Cont.}
% 	%	\caption[Radial profile of the scale height for the UGC 09977 galaxy]{Radial profile of the scale height for the UGC 09977 galaxy. \textit{Top panel}: The applied box model of the galaxy. Each bin has a size of $8 \times 4$ pixels. \textit{Bottom panel}: The distribution of the scale height along the radius. Error bars represent statistical uncertainties from fits.}
% 	\label{fig:radial_profile_of_hz_sech_part3}
% \end{figure*}
\begin{figure}
	\vspace{-0.5cm}
	\centering
	\includegraphics[width=1\linewidth]{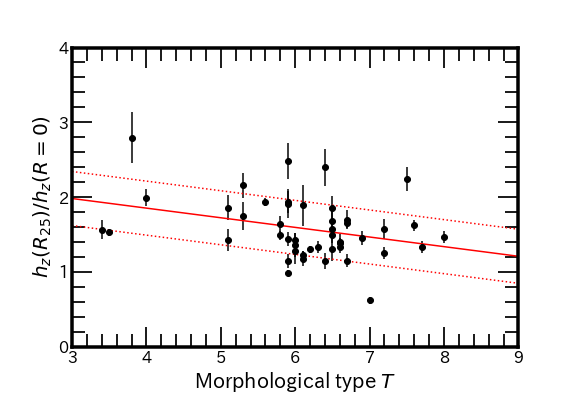}
	\caption{Ratio of the scale heights at $R_{25}$ and $R=0$ against the morphological type T. Error bars are uncertainties issued from statistical errors.}
	\label{fig:r25_r0_vs_T}
\end{figure}

\subsection{Correlation with galaxy global properties}\label{sec:correlations}
We discuss in this section possible correlations that exist between the measured scale height with other galaxy properties. 
%\subsubsection{Scale height vs radial-to-vertical scale length ratio $q_R$}

\subsubsection{Scale height and scale length}
\begin{figure}
	%	\centering
	%\begin{minipage}{0.9\textwidth}
	%\vspace{-0.5cm}
	\centering
	\includegraphics[width=\columnwidth]{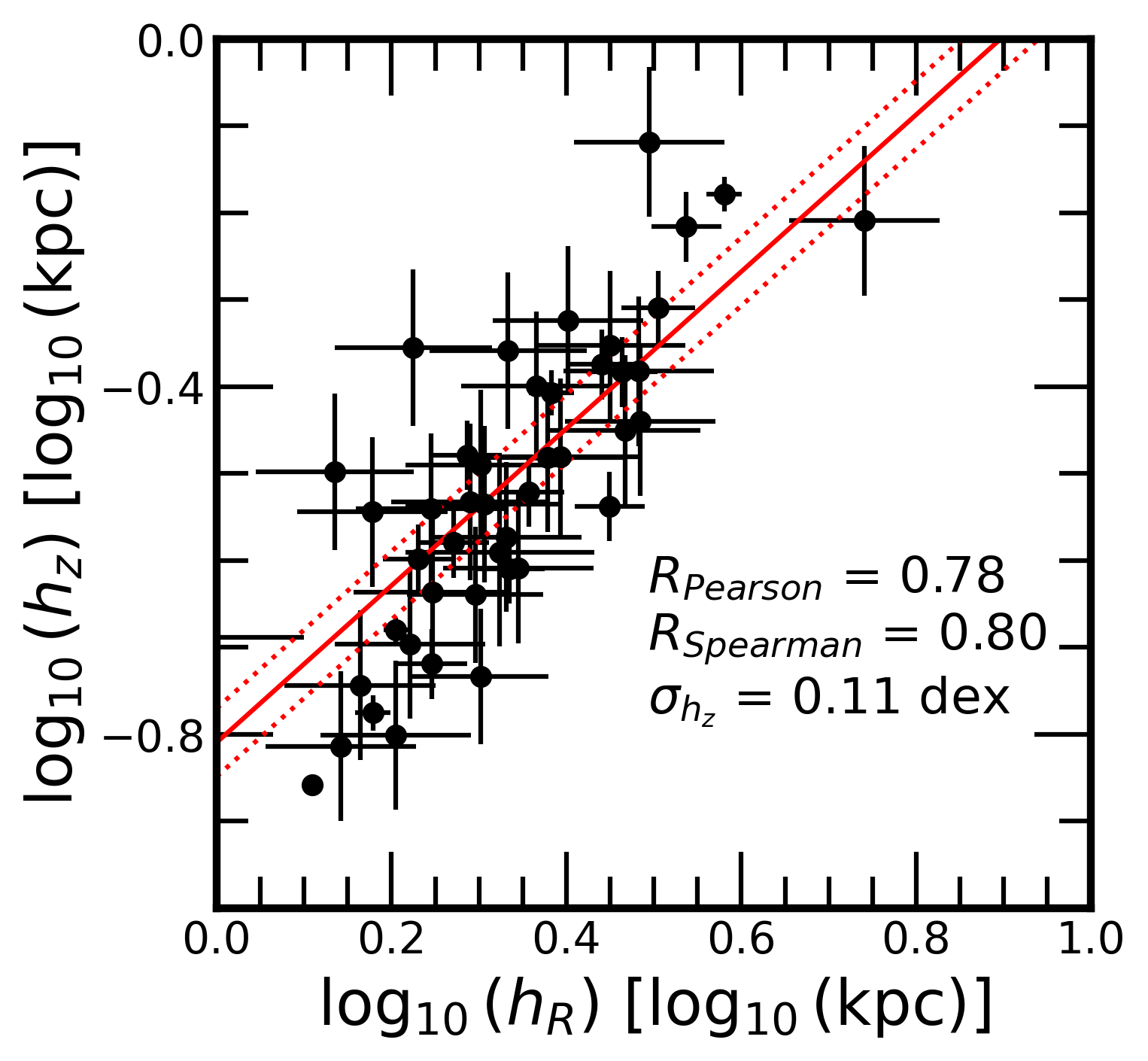}
	\caption{Relationship between scale height $h_z$ and scale length $h_R$ for single-component fit results. The observed tight correlation can be used as an empirical relation for scale height estimation.}
	\label{fig:loghz_vs_hR}
	%\end{figure}
\end{figure}

We find a good correlation between the single-component scale height estimated using \textsc{Galfit} and scale length as seen in Figure \ref{fig:loghz_vs_hR}. The best-fit linear model obtained (with a p-value $p<0.001$, $R_\text{Pearson}= 0.78$, $R_\text{Spearman}= 0.80$  and $ \sigma_{h_z}=0.11 $ dex) is:
\begin{equation}\label{eq:hz_vs_hR}
	\log_{10}{(h_z/\text{kpc})} = 0.90 \log_{10}{(h_R/\text{kpc})} - 0.81 \pm 0.04  
\end{equation}
This relationship can be used to estimate the scale height using the scale length for spiral galaxies with morphological types ranging from Sb to Sdm. 
%It is also interesting to see if such a relationship exists for the thin and thick disks. However, we leave such an investigation as future work as estimating the scale length of multiple component disks is beyond the scope of this study.

\begin{figure*}
	%	\vspace{-2cm}
	%	\hspace{-2.6cm}
	\centering
	\begin{minipage}{0.475\textwidth}
		%\begin{figure}
		%\vspace{-0.5cm}
		\centering
		\includegraphics[width=1\textwidth]{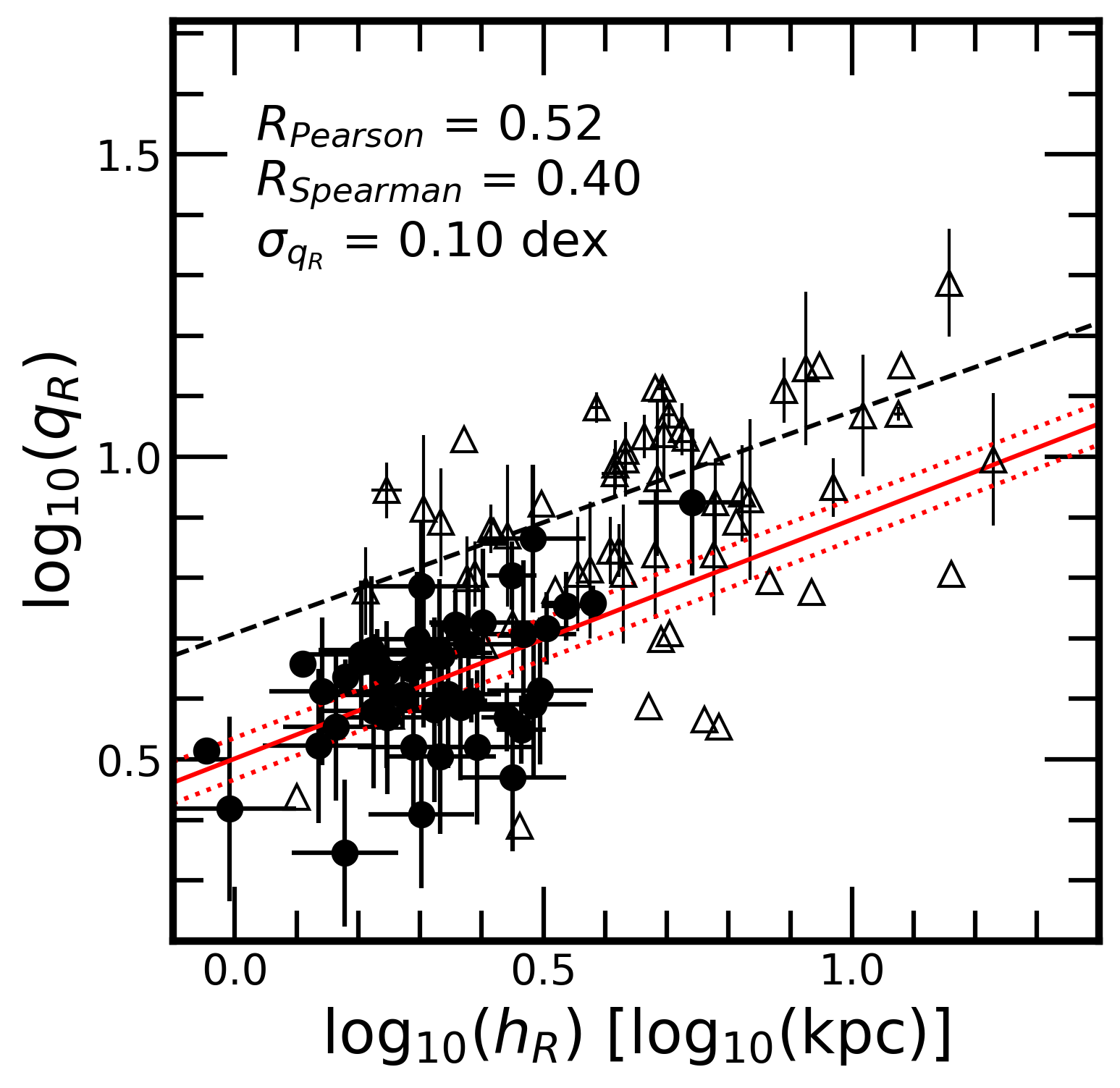}
	\end{minipage}
	\hfill
	\begin{minipage}{0.475\textwidth}
		%\begin{figure}
		%\vspace{-0.5cm}
		\centering
		\includegraphics[width=1\textwidth]{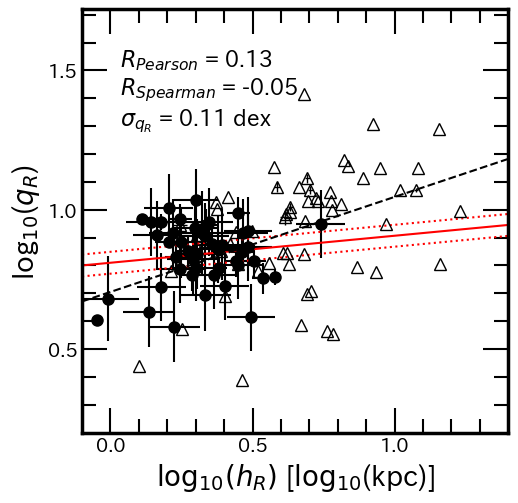}
	\end{minipage}
	\caption{The disk oblateness against the scale length from our results (closed circles) and data from \citet{bershady2010diskmass} for Sb $ - $ Sdm galaxies in I-, H-, K- bands (open triangles). The left panel shows uncorrected data for inclination. The right panel has corrected data points.}
	\label{fig:qR_vs_hR}
\end{figure*}

\cite{bershady2010diskmass} also found a relation between the radial-to-vertical scale length ratio and the scale length using I-, H- and K-band imaging. Some of their data points neglected inclination correction as these galaxies had $ i\geq87^\circ $ (see Section \ref{sec:litreview}). Our measured scale heights were corrected for inclination using Equation \ref{eq:inclcorrection}. We also correct the scale height collected from the literature for comparison. The correlation between the disk oblateness and the disk scale length is shown in Figure \ref{fig:qR_vs_hR}, on the left panel, the scale heights are not corrected for inclination and the corrected values are plotted on the right panel. This figure shows that we are able to reproduce the relationship in \citet{bershady2010diskmass} with $ R_\text{Pearson} = 0.52 $, $ R_\text{Spearman} = 0.40 $ and $p<0.001$ when the scale height is not corrected for inclination. However, this correlation disappears when the scale height is corrected for inclination ($ R_{\text{Pearson}}=0.13 $, $ R_{\text{Spearman}}=-0.05 $). This implies that the q$_{R}$-h$_{h}$ correlation is more susceptible to uncertainties compared to the h$_{z}$-h$_{R}$ correlation shown in Figure \ref{fig:loghz_vs_hR}.
%The relation we found for the corrected $q_R$ is:   
%To cope with this issue, we corrected the scale heights using relation \ref{eq:inclcorrection} and indeed, no flagrant change was to report. In the right panel of Figure \ref{fig:qR_vs_hR}, we compare the correlation we found (closed circles) with theirs for Sb to Sdm type galaxies (open triangles). Unfortunately, our relationship only returned $ R_{\text{Pearson}}=0.13 $ and $ \sigma_{q_R} = 0.11 $ dex. This is mainly because of the inclination correction on our scale height results. In fact, the found correlation is a lot better without the correction ($ R_{\text{Pearson}}=0.52 $ and $ \sigma_{q_R} = 0.10 $) as shown in the left panel of Figure \ref{fig:qR_vs_hR}. 
%\begin{equation}
%	\log_{10}{(q_R)} = 0.10 \log_{10}{(h_R/\text{kpc})} + 0.81 \pm 0.04  
%\end{equation}
%against their $ R_{\text{Pearson}}=0.41 $ and $ \sigma_{q_R} = 0.19 $ dex (for Sb $ - $ Sdm galaxies after inclination correction)


\subsubsection{Scale height and the apparent radius $ R_{25} $}
\begin{figure*}
	%\vspace{-4cm}
	\centering
	\begin{minipage}{0.475\textwidth}
		%\vspace{-0.5cm}
		\centering
		\includegraphics[width=\textwidth]{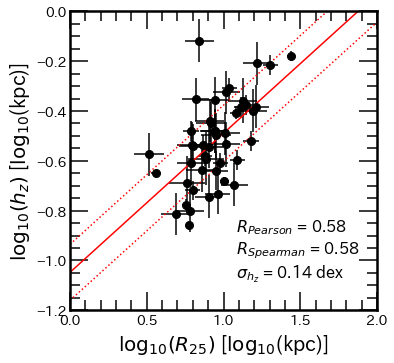}
	\end{minipage}
	\begin{minipage}{0.475\textwidth}
		%\vspace{-0.5cm}
		\centering
		\includegraphics[width=\textwidth]{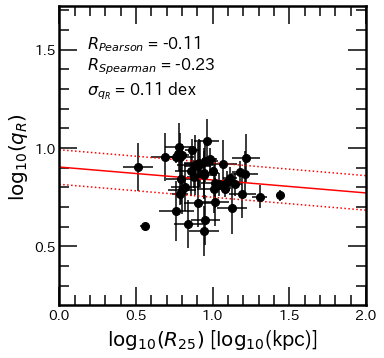}
	\end{minipage}
	\caption{\textit{Left panel:} Correlation between single-component scale height $h_z$ and the uncorrected apparent radius $R_{25}$. \textit{Right Panel:} Oblateness and $R_{25}$.}
	\label{fig:loghz_vs_r25}
\end{figure*}

\begin{figure*}
	%	\hspace{-2cm}
	\centering
	\begin{minipage}{0.475\textwidth}
		%\vspace{-0.5cm}
		\centering
		\includegraphics[width=1\textwidth]{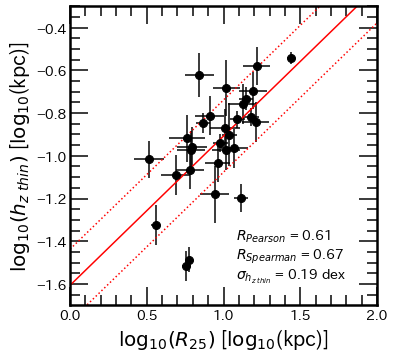}
		%\end{figure}
	\end{minipage}
	\hfill
	\begin{minipage}{0.475\textwidth}
		%\begin{figure}
		%\vspace{-0.5cm}
		\centering
		\includegraphics[width=\textwidth]{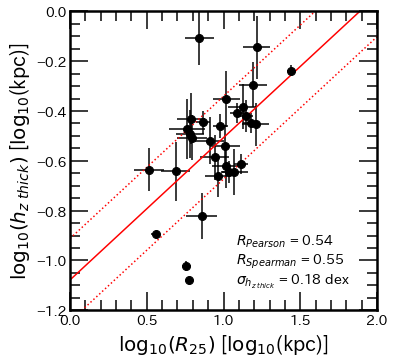}
	\end{minipage}
	\caption{\textit{Left panel}: Thin disk scale height as a function the apparent radius $R_{25}$. \textit{Right panel}: Relationship between thick disk scale height and $R_{25}$.}
	\label{fig:log2comphz_vs_r25}
\end{figure*}

Figure \ref{fig:loghz_vs_r25} shows the correlation between $h_{z}$ and $ R_{25} $  for the single-component disk analysis. 
The resulting relation between the scale height and $ R_{25} $ (left panel) is:
\begin{equation}
	\log_{10}{(h_z/\text{kpc})} = 0.56 \log_{10}{(R_{25}/\text{kpc})} - 1.05 \pm 0.11
\end{equation}
This relationship could be used to estimate $h_z$ as $R_{25}$ are widely available for most galaxies ($ p<0.001 $). The disk oblateness $q$ as a function of the apparent radius is shown in the right panel of Figure \ref{fig:loghz_vs_r25} with weak correlation coefficients of $ R_{\text{Pearson}} = -0.11 $ and $ R_{\text{Spearman}} = -0.23 $ as well as a p-value of 0.47. We also report a correlation between $h_{z \ \text{thin}}$ and $R_{25}$ but also between $h_{z \ \text{thick}}$ and $R_{25}$ (Figure \ref{fig:log2comphz_vs_r25}). While the correlations are comparable with the one obtained from single-component $h_z$, the relation obtained from the thick disk is slightly weaker and the one from the thin disk is significantly better (for the thick disk $R_{\text{Pearson}}=0.54; \ R_{\text{Spearman}}=0.55; \ p=0.002; \ \sigma_{h_{z \ \text{thin}}} = 0.18 $ dex and for the thin disk $R_{\text{Pearson}}=0.61; \ R_{\text{Spearman}}=0.67; \ p<0.001; \ \sigma_{h_{z \ \text{thin}}} = 0.18 $ dex). This could be due to the fact that both the thin disk and $ R_{25} $ are at the edge of star-forming regions. The best-fit linear models are:
\begin{align}
	\log_{10}{(h_{z \ \text{thin}}/\text{kpc})} &= 0.69 \log_{10}{(R_{25}/\text{kpc})} - 1.59 \pm 0.16 \\
	\log_{10}{(h_{z \ \text{thick}}/\text{kpc})} &= 0.56 \log_{10}{(R_{25}/\text{kpc})} - 1.07 \pm 0.16
\end{align}


\subsubsection{Scale height and $ V_\text{max} $ }

\begin{figure*}
	%	\hspace{-2cm}
	\centering
	%	\begin{minipage}{0.32\textwidth}
		%		%\vspace{-0.5cm}
		%		\centering
		%		\includegraphics[width=\textwidth]{loghz_vs_vmax.png}
		%		%\end{figure}
		%	\end{minipage}
	%	\hfill
	%	\begin{minipage}{0.475\textwidth}
		%		%\begin{figure}
		%		%\vspace{-0.5cm}
		%		\centering
		%		\includegraphics[width=\textwidth]{loghzthin_thick_vs_vmax_var_weights.png}
		%	\end{minipage}
	%	\caption[Relationship between scale height and maximum velocity]{Scale height as a function of the maximum velocity $V_\text{max}$. \textit{Left panel}: Correlation obtained from single-component fits. \textit{Right panel}: Relationship for two-component disks: the thin disk (closed circles) and the thick disk (open circles).}
	%	\label{fig:loghz_vs_vmax}
	\begin{minipage}{0.475\textwidth}
		%\begin{figure}
		%\vspace{-0.5cm}
		\centering
		\includegraphics[width=\textwidth]{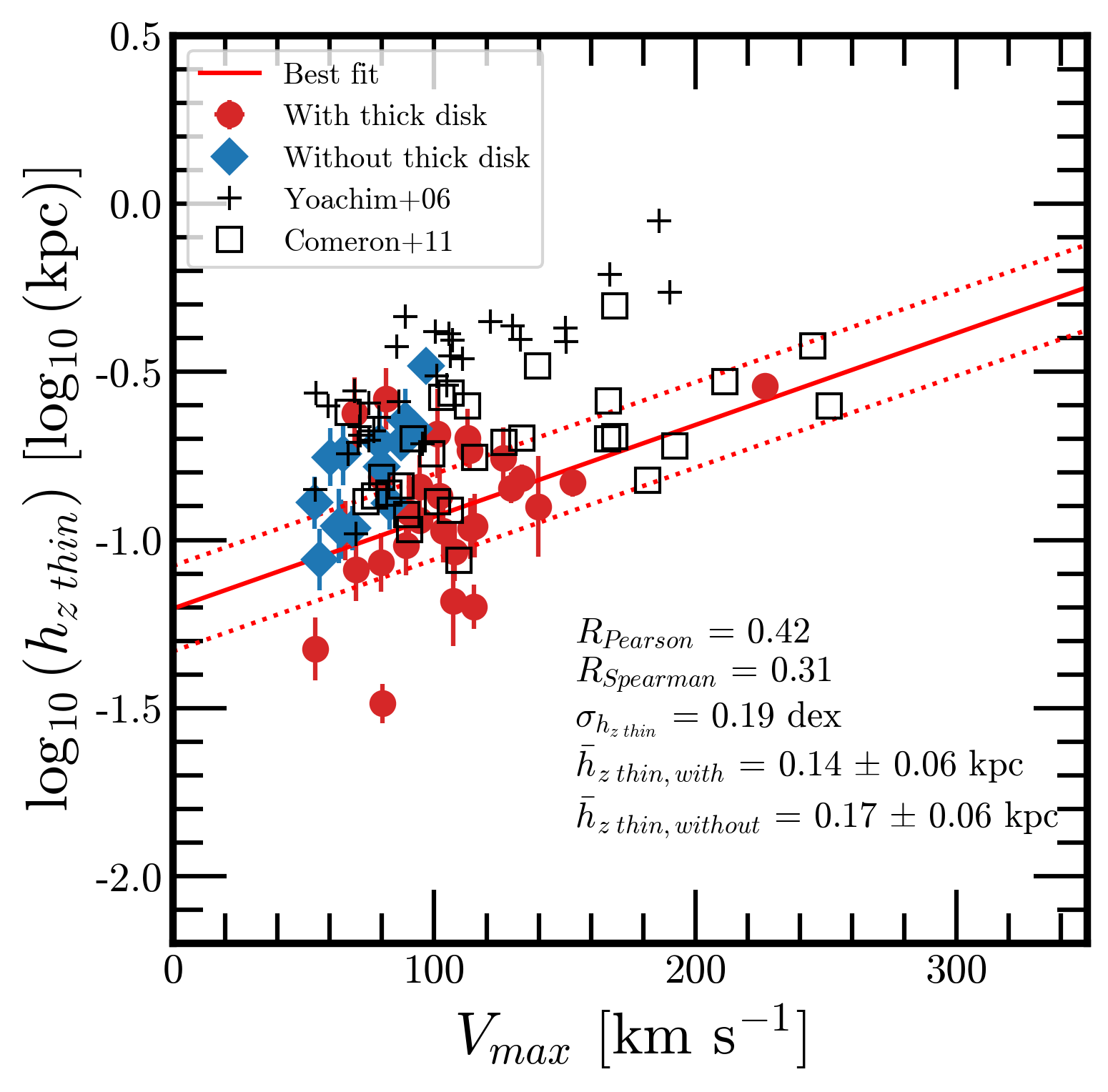}
	\end{minipage}
	\hfill
	\begin{minipage}{0.475\textwidth}
		%\begin{figure}
		%\vspace{-0.5cm}
		\centering
		\includegraphics[width=\textwidth]{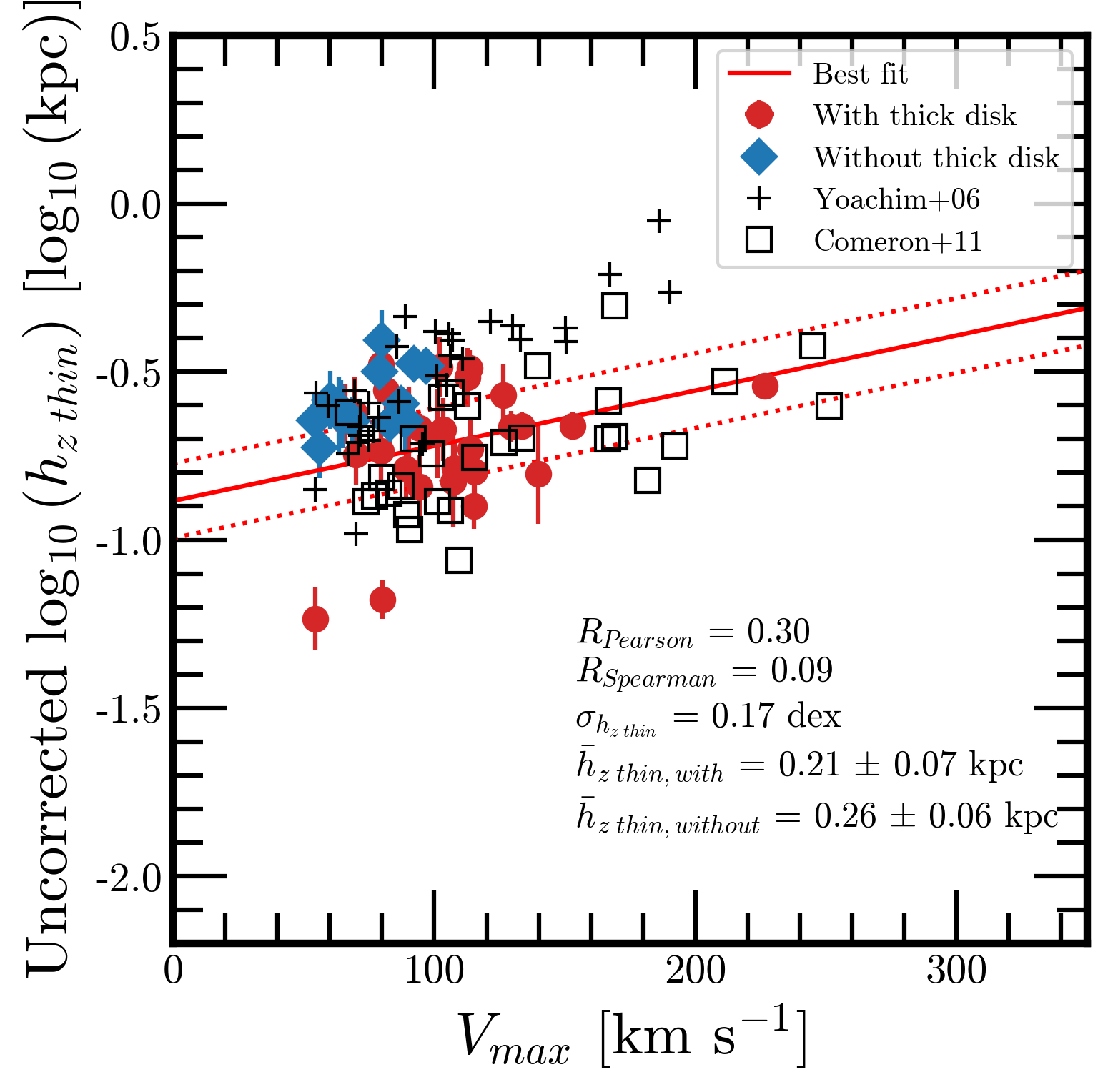}
	\end{minipage}
	\caption{Scale height of the thin disk as a function of the maximum velocity $V_\text{max}$. Red solid round and blue diamond points represent galaxies with and without a thick disk respectively. Note that both dust-obscured and unobscured data from \citetalias{2006AJ....131..226Y} have been plotted. \textit{Left panel}: Correlation obtained from inclination-corrected scale heights. \textit{Right panel}: Relationship for the uncorrected scale heights.}
	\label{fig:loghz_vs_vmax}
\end{figure*}

\begin{figure}
	\centering
	\includegraphics[width=1\linewidth]{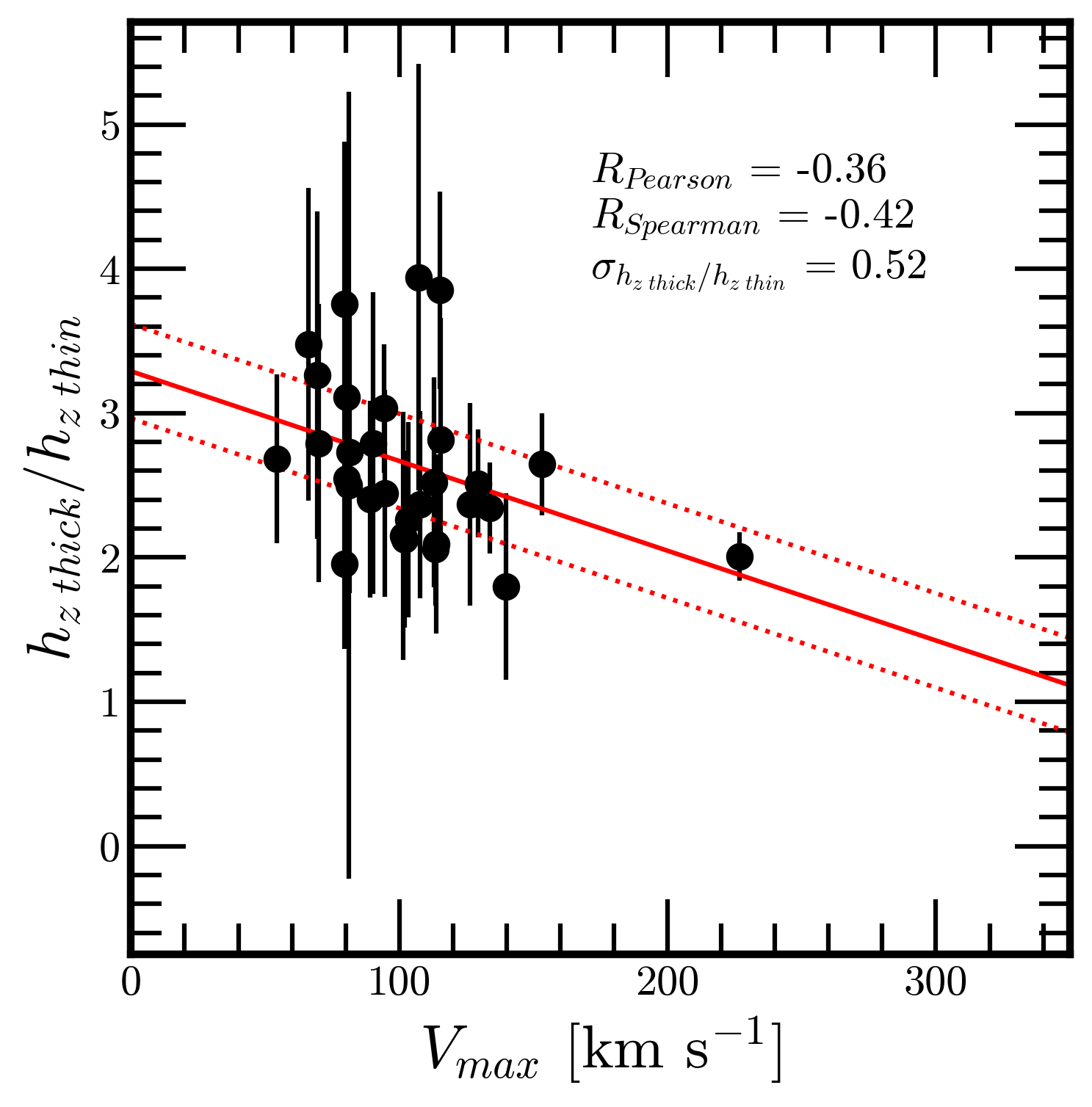}
	\caption{Scale height ratio and maximum velocity.}
	\label{fig:loghzratio_vs_vmax}
\end{figure}

In the left panel Figure \ref {fig:loghz_vs_vmax}, we show the correlation between the thin disk scale heights and the maximum velocity $V_\text{max}$ retrieved from HyperLEDA. This figure shows that the scale heights increase with $V_\text{max}$, although the scatter tends to be greater for lower maximum velocity $V_\text{max} \leq 100$ km s$^{-1}$. The correlation have a Pearson correlation coefficient of 0.42, a Spearman coefficient of 0.31 and a p-value of 0.03 with an overall scatter of 0.19 dex. This trend is consistent with the previous findings of \citetalias{2006AJ....131..226Y} and \citet{2011ApJ...741...28C}. However, a majority of our scale heights are consistently lower than the ones from \citetalias{2006AJ....131..226Y}. Nonetheless, our uncorrected-for-inclination values are relatively closer to their $h_{z \ \text{thin}}$ (right panel) an observation that is previously discussed in Section \ref{sec:res2comp}. We report a weak correlation between the uncorrected thin disk scale heights and $V_\text{max}$ ($R_\text{Pearson} = 0.30; \ R=\text{Spearman} = 0.09; \ p= 0.12; \ \sigma_{h_z \ \text{thin}} = 0.17$ dex. Galaxies with bad thick disk results (considered to be without thick disk) are observed to have $V_\text{max} < 100 \ \text{km s}^{-1}$. This agrees with the observations of \citet{2018A&A...610A...5C} who found that all the 17 galaxies without a thick disk in their sample of 141 galaxies have a circular velocity below $120 \ \text{km s}^{-1}$ accounting for the gas and a possible third stellar component. However, we also find a moderate anti-correlation between the scale height ratio and the maximum velocity (Figure \ref{fig:loghzratio_vs_vmax}) with $R_\text{Pearson} = -0.36; \ R_\text{Spearman} = -0.42; \ p=0.05$ that shows that thick disks are more prominent in lower mass, two-component disk galaxies.
 
%Galaxies with bad thick disk results (considered to be without thick disk) are observed to be thicker than lower mass systems with a thick disk. 
%The observed trend seem to contradict the findings of \citetalias{2006AJ....131..226Y} and \citet{2011ApJ...741...28C} that suggest the thick disks of lower velocity galaxies more readily detectable due to their prominence.

%In Figure \ref {fig:loghz_vs_vmax}, we show the correlation between the scale heights of thin disk and the maximum velocity $V_\text{max}$ retrieved from HyperLEDA for single-component $h_z$ from \textsc{Galfit} (left panel) and $h_{z \ \text{thin}}$ (right panel). This figure shows that the scale heights increase with $V_\text{max}$, although the scatter tends to be greater for lower maximum velocity $V_\text{max} \leq 200$ km s$^{-1}$. These correlations have a Pearson correlation coefficient of 0.35 and 0.39, a Spearman correlation coefficient of 0.30 and 0.32 and a p-value of 0.02 and 0.03 with a scatter ranging of 0.16 to 0.22 dex respectively. 
The corresponding relations are:
%The correlation from the scale height ratio gave $R_\text{Pearson} = -0.31$ with a scatter of 20\%. The thick-to-thin scale height ratio also decreases with increasing $V_\text{max}$ (Figure \ref {fig:hzratio_vs_vmax}). 
\begin{align}
%	\log_{10}{(h_{z}/\text{kpc})} &= 0.10 \ \frac{V_\text{max}}{100 \ \text{km s}^{-1}} - 0.72 \pm 0.09 \\
	\log_{10}{(h_{z \ \text{thin}}/\text{kpc})} &= 0.27 \ \frac{V_\text{max}}{100 \ \text{km s}^{-1}} - 1.20 \pm 0.13 \\
	\frac{h_{z \ \text{thick}}}{h_{z \ \text{thin}}} 
	&= -0.62 \ \frac{V_\text{max}}{100 \ \text{km s}^{-1}} + 3.29 \pm 0.33	
\end{align}
\subsubsection{Scale height and other global and photometric properties}\label{sec:otherCorrelations}
\begin{figure*}
	%\hspace{-2cm}
	\centering
	\begin{minipage}{0.475\textwidth}
		\centering
		\includegraphics[width=1\textwidth]{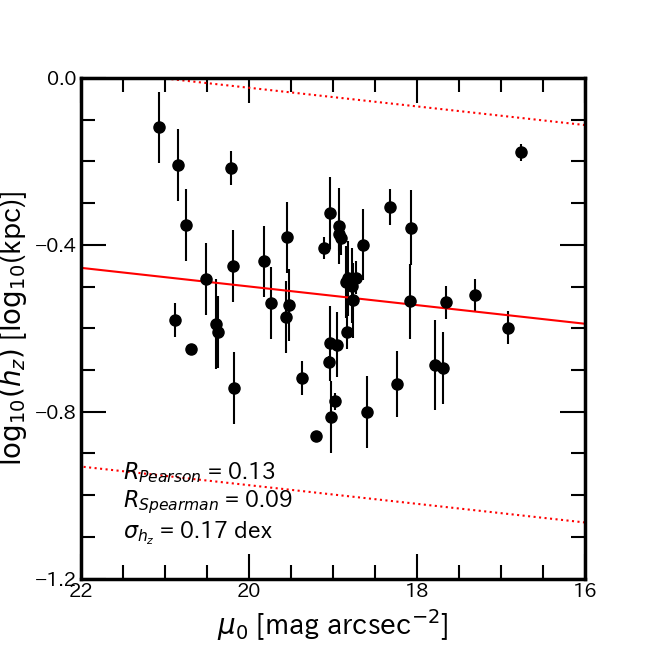}
	\end{minipage}
	\hfill
	\begin{minipage}{0.475\textwidth}
		\centering
		\includegraphics[width=1\textwidth]{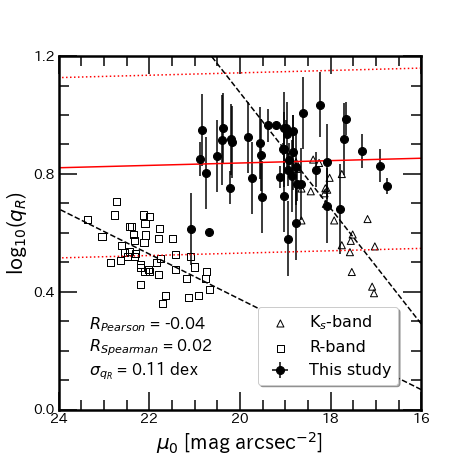}
	\end{minipage}
	\caption{\textit{Left Panel:} Single-component scale heights as a function of the face-on central surface brightness. \textit{Right Panel:} Single-component radial-to-vertical scale length ratio against the face-on central surface brightness. Our results are represented by the dark circles. Square and triangle symbols are respectively the B- and K$ _\text{s} $-band data from \cite{2003EAS....10..121Z}. Dashed dark lines indicate linear fits from these data points.}
	\label{fig:loghz_vs_mu0}
	
\end{figure*}

%\subsubsection{Scale height and central surface brightness $\mu_{3.6\mu m} (0, 0)$}
Left panel of Figure \ref{fig:loghz_vs_mu0} shows the one-component scale height $h_z$ from \textsc{Galfit} as a function of the central surface brightness $\mu_{3.6\mu m} (0, 0)$. We report a weak correlation with correlation coefficients of $R_\text{Pearson} = 0.13$ and $R_\text{Spearman} = 0.09$, a p-value of 0.37 and a scatter $\sigma_{h_z}=0.17 $ dex. For consistency with previous studies, we also test our radial-to-vertical scale length ratio $q_R = h_R/h_z$ against $\mu_{3.6\mu m} (0, 0)$. In the right panel of Figure \ref{fig:loghz_vs_mu0}, we present the correlation we find along with scaling relation obtained by \cite{2003EAS....10..121Z} using R- and K$_\text{s}$- band data. Both their samples (square symbols indicating R-band data points and triangles representing K$_\text{s}$-band elements) returned good correlations, however, our data shows no such correlation (back full circles) yielding a weak interdependence ($R_\text{Pearson} = -0.04$, $R_\text{Spearman} = 0.02$, $ p=0.80 $, $\sigma_{q_R}=0.11 $ dex). The reason why the correlation is weak in the 3.6 $\mu$m-band is unclear. However, unlike our direct model-fitting methods, they estimated the scale height from the scale length and using the observed axis ratio. Additionally, if such a relation exists, it is likely wavelength-dependent as the correlation from \cite{2003EAS....10..121Z} seem to be different in the R- and K$_\text{s}$- band.
%%Although the existence of such interdependence is expected,
%

We also tested our scale height results as well as $ h_z $ ratio against the total stellar mass $M_\star$ and the morphological type T (both values were issued from the S$ ^4 $G archive). However, the relations that we find have $ p>0.05 $ and thus have been rejected. 

\subsubsection{Correlation comparison}\label{sec:CorrelationComp}
\begin{figure*}
	%	\hspace{-2cm}
	%\centering
	\begin{minipage}{0.32\textwidth}
		\centering
		\includegraphics[width=1\textwidth]{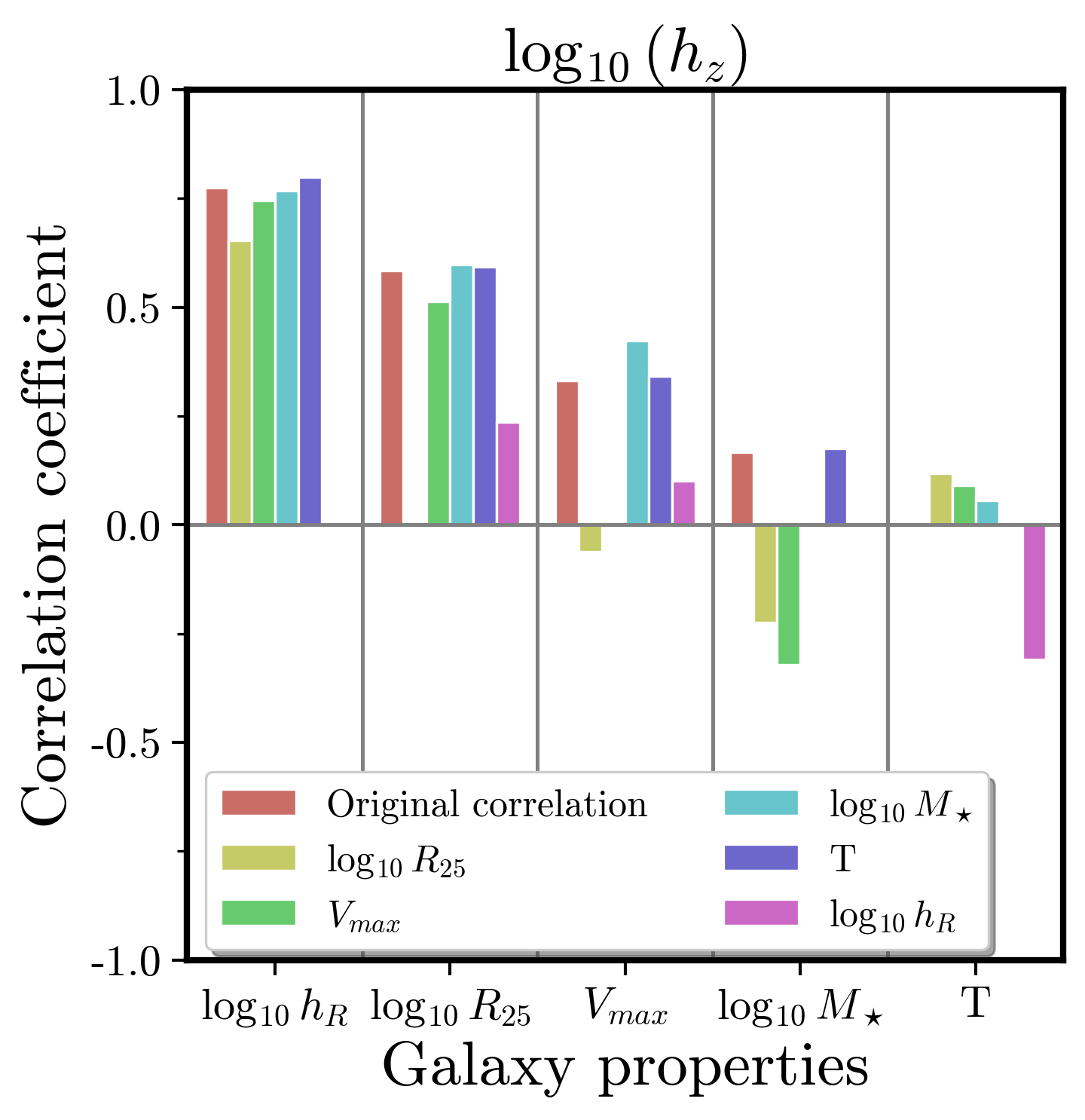}
	\end{minipage}
	\hfill
	\begin{minipage}{0.32\textwidth}
		\centering
		\includegraphics[width=1\textwidth]{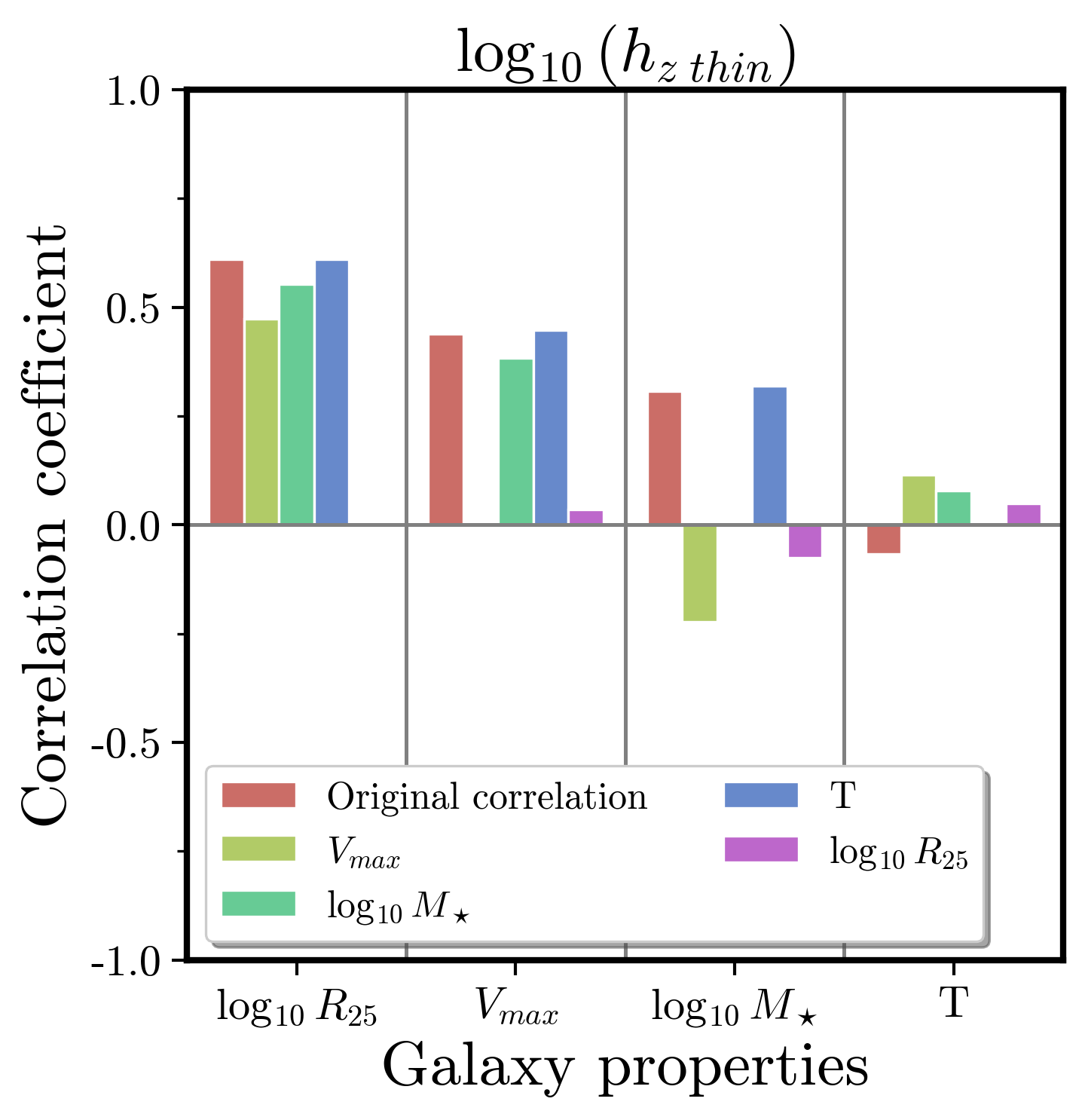}
	\end{minipage}
	\hfill
	\begin{minipage}{0.32\textwidth}
		\centering
		\includegraphics[width=1\textwidth]{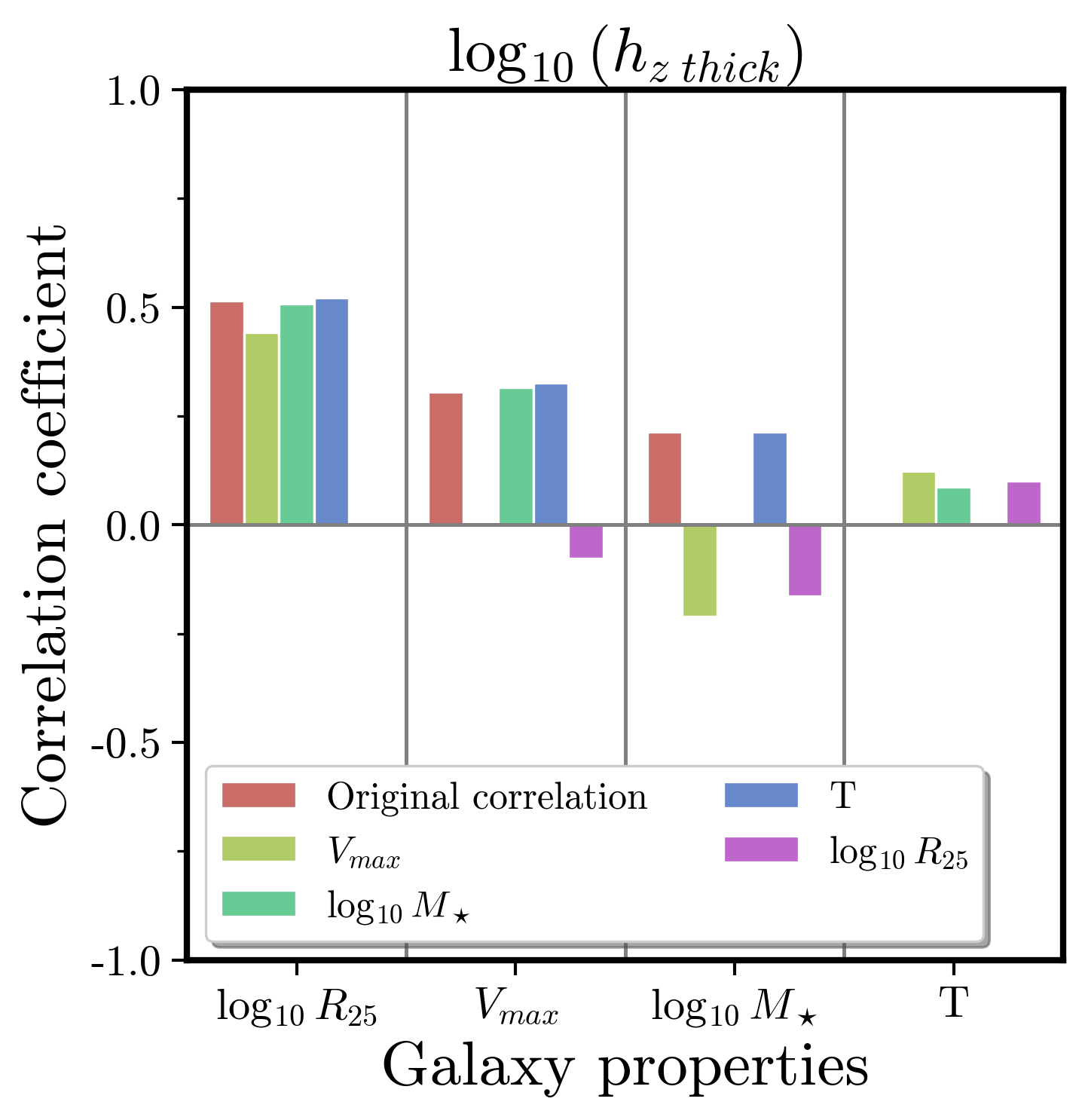}
	\end{minipage}
	\caption{Partial correlations between scale heights and galaxy properties with control parameters (color bars). $h_R$ represents the radial scale length,  $R_{25}$ the apparent radius, $V_\text{max}$ the maximum velocity. $M_\star$ the current stellar mass and T the morphological type. Red bar shows the original correlation coefficient. \textit{Left Panel:} Correlations for single-component scale height. \textit{Middle Panel:} Relationships for thin disk scale height. \textit{Right Panel:} Thick disk scale height and galaxy properties.}
	\label{fig:partialcorrelation}
\end{figure*}

To probe the a fundamental relation amongst the different correlations, we perform partial correlations for single and two-component scale heights as presented in Figure \ref{fig:partialcorrelation}. We observe that the most intrinsic correlation is the $h_z - h_R$ relationship. The correlation between the scale height and apparent radius is the second-best relationship that we found. However, Figure \ref{fig:partialcorrelation} shows that this correlation stems from a more intrinsic relation with $h_R$. The $h_z - V_\text{max}$ relation is the third-most intrinsic correlation we found. This relationship is dependent on $h_R$ and $R_{25}$ as observed in the figure. This is reasonable due to the existence of the intrinsic $h_z - h_R$ correlation as well as the size-velocity relation (e.g. \citealt{1998MNRAS.295..319M}; \citealt{2021SHPSA..88..220M}).

\section{Summary} \label{chap:summary}
We have measured the stellar disk thickness of 46 edge-on galaxies selected from the Spitzer Survey of Stellar Structure in Galaxies S$ ^4 $G survey by fitting the vertical profile using simple models. Near-infrared images are known to trace the bulk of the stellar mass and are less affected by dust obstruction. The effect of inclination is carefully corrected in all measurements using synthetic images from the GALMER simulations. We use 1D, 2D, and 3D techniques to measure the disk thickness and coherently compare the performance of these three methods.
%We compared the stellar disk scale height measured using one, two and three-dimensional methods. We also investigated possible correlations between the disk vertical scale have with other galaxy properties and explored the thin and thick disk models.
Our findings are the following: 
\begin{itemize}
 \item After comparing the uncertainties of the results from the different methods, we found that the 3D Imfit and 2D \textsc{Galfit} EdgeOnDisk models have the lowest uncertainties. The single disk component fits showed that our inclination correction 1D measurements were consistent with 3D models. 

 \item Our two-component fits reveal that about two-thirds of our sample showed evidence of a thick disk. Yet, none of our galaxies have the traditional thick disk scale height of $ h_{z \ \text{thick}} \geq 1 \text{kpc} $. This could be the effect of our sample size or the wavelength band that is used. Most of the disks are also vertically symmetrical. We also find that thick disks are 2.65 times larger than thin disks on average. Our analyses on the scale height at specific radii of galaxy disks also prove the existence of disk flaring which is consistent with previous studies. The flaring amplitude is more significant for early-type spirals. 

 \item We also found correlations between the vertical scale height with other galaxy properties as follows:

\begin{enumerate}
	\item The single-component scale height grows with increasing scale length. However, the radial-to-vertical scale length ratio seems to be constant.
	\item Scale heights from both one and two-component models rise with apparent radius $R_{25}$.
%	\item Galaxies of higher current total stellar mass ($M_\star \geq 10^9 M_\odot$) tend to have larger two-component scale heights but lower thick-to-thin scale height ratio. 
%	\item Thicker disks seem to have more active star formation in them. While this may be true, galaxies with a lower scale height ratio have a bigger star formation rate.
	\item One-component and thin disk $ h_z $ tend to increase with the maximum velocity. 
%	\item Disks with higher maximum velocity are likely to be thicker while having a less prominent thick disk.
%	\item Later-type spirals have higher scale height ratios.
\end{enumerate}
\end{itemize}
After performing partial correlation tests, we find that the relation between the scale height and scale length is an intrinsic relationship. This relation could be used for single-component scale height estimations. The relation between scale height and $R_{25}$ can also be used for two-component thickness estimation.

\section*{Acknowledgements}
NR acknowledges financial support from the IAU GA, Astronomy in Africa Scholarship.
KMD and TSG thanks the support of the Serrapilheira Institute (grant Serra-1709-17357) as well as that of the Brazilian National Research Council (CNPq 308584/2022-8) and of the Rio de Janeiro Research Foundation (FAPARJ grant E-32/200.952/2022), Brazil.
The financial assistance of the South African Radio Astronomy Observatory (SARAO) towards this research is hereby acknowledged (\url{www.sarao.ac.za}).
%The Acknowledgements section is not numbered. Here you can thank helpful
%colleagues, acknowledge funding agencies, telescopes and facilities used etc.
%Try to keep it short.

%%%%%%%%%%%%%%%%%%%%%%%%%%%%%%%%%%%%%%%%%%%%%%%%%%
\section*{Data Availability}

The data used in this work are publicly available as part of the S4G project (\url{https://irsa.ipac.caltech.edu/data/SPITZER/S4G/}). Derived properties and the scripts used for the analysis will be made available upon reasonable request to the author.

\bibliography{sample631}{}
\bibliographystyle{aasjournal}

%% This command is needed to show the entire author+affiliation list when
%% the collaboration and author truncation commands are used.  It has to
%% go at the end of the manuscript.
%\allauthors

%% Include this line if you are using the \added, \replaced, \deleted
%% commands to see a summary list of all changes at the end of the article.
%\listofchanges

\end{document}